.

# Faint recombination lines in Galactic PNe with [WC] nucleus ⋆

Jorge García-Rojas[1], Miriam Peña[1], and Antonio Peimbert[1]

Instituto de Astronomía, Universidad Nacional Autónoma de México; Apdo. P ostal 70264; México D. F., 04510; México

**ABSTRACT**

*Aims.* We present spatially resolved high-resolution spectrophotometric data for the planetary nebulae PB 8, NGC 2867, and PB 6. We have analyzed two knots in NGC 2867 and PB 6 and one in PB 8. The three nebulae are ionized by [WC] type nuclei: early [WO] for PB 6 and NGC 2867 and [WC 5-6] in the case of PB 8. Our aim is to study the behavior of the abundance discrepancy problem (ADF) in this type of planetary nebulae.
*Methods.* We measured a large number of optical recombination (ORL) and collisionally excited lines (CEL), from different ionization stages (many more than in any previous work), thus, we were able to derive physical conditions from many different diagnostic procedures. We determined ionic abundances from the available collisionally excited lines and recombination lines. Based on both sets of ionic abundances, we derived total chemical abundances in the nebulae using suitable ionization correction factors.
*Results.* From CELs, we have found abundances typical of Galactic disk planetary nebulae. Moderate ADF($O^{++}$) were found for PB 8 (2.57) and NGC 2867 (1.63). For NGC 2867, abundances from ORLs are higher but still consistent with Galactic disk planetary nebulae. On the contrary, PB 8 presents a very high O/H ratio from ORLs. A high C/O was obtained from ORLs for NGC 2867; this ratio is similar to C/O obtained from CELs and with the chemical composition of the wind of the central star, indicating that there was no further C-enrichment in the star, relative to O, after the nebular material ejection. On the contrary, we found C/O<1 in PB 8. Interestingly, we obtain $(C/O)_{ORLs}/(C/O)_{CELs} < 1$ in PB 8 and NGC 2867; this added to the similarity between the heliocentric velocities measured in [O III] and O II lines for our three objects, argue against the presence of H-deficient metal-rich knots coming from a late thermal pulse event.

## 1. Introduction

Great efforts have been devoted in recent years to analyze the discrepancies between the heavy element abundances derived from collisionally excited lines (CELs) and those derived from optical recombination lines (ORLs) in diffuse nebulae. Such discrepancies are commonly quantified using the so-called abundance discrepancy factor (ADF), which is defined as:

$$\text{ADF}(X^{i+}) = (X^{i+}/H^+)_{\text{ORLs}}/(X^{i+}/H^+)_{\text{CELs}}. \qquad (1)$$

Typical values for the ADF are 1.4 – 2.8 in HII regions (see García-Rojas & Esteban, 2007, and references therein), and 1.6 – 3 in planetary nebulae (PNe), but much larger values have been reported for some PNe, e.g., ADFs of 10 for NGC 6153 and 70 for Hf 2-2 (see Liu et al., 2006, and references therein)

The exact causes for these discrepancies are still unknown. One possibility is that such discrepancy could be due to thermal inhomogeneities inside the nebula; such inhomogeneities are characterized by the parameter $t^2$ as introduced by Peimbert (1967). Thermal inhomogeneities could be produced by extra heating mechanisms, besides the ionizing stellar photons, inside a chemically homogeneous nebula (see Peimbert & Peimbert, 2006, and references therein). Alternatively they could be due to chemical inhomogeneities inside the nebula; these have been modeled as very enriched knots inside the nebula, where the electron temperature is much lower than in the diffuse ionized gas, in such cold knots the ORLs would be much enhanced while the CELs would be emitted in the diffuse hot gas (see Liu, 2006, and references therein).

Among the few hundreds of PNe with studied stellar continuum, only 15% of them are ionized by stars presenting Wolf-Rayet features. These stars show in their spectra prominent wide emission lines of C, O, and He due to an intense wind characterized by a large mass-loss rate. All these stars belong to the C-sequence of Wolf-Rayet stars (hereafter [WC] stars), mostly of the spectral types [WC 2–4] and [WC 8–11], with very few objects in the intermediate classes (Tylenda et al., 1993; Crowther et al., 1998; Acker & Neiner, 2003).

The stellar atmospheres are almost pure helium and carbon (e.g. Hamann, 1997). Several scenarios to produce such H-deficient low-mass stars have been proposed (e.g. Blöcker, 2001; Herwig, 2001). In addition, Górny & Tylenda (2000), De Marco (2002), and Medina et al. (2006) have provided arguments for the existence of the evolutionary sequence proposed by Acker et al. (1996) and Hamann (1997): [WC]-late → [WC]-early stars, ending with the PG 1159 type stars. However, this proposition is still under debate (Hamann et al., 2005; Todt et al., 2008).

It has been shown that the mechanical energy of the stellar wind deposited on the nebular shell around a [WC] star (WRPN) produces higher expansion velocity and higher turbulence in the shell than in non-WRPNe (Medina et al., 2006; Gesicki et al., 2003). So far no other effects of the stellar wind have been found. In particular, the chemical abundances of WRPNe seem to follow the same patterns as non-WRPNe, with no particular enrichment (Gorny & Stasińska, 1995; Peña et al., 2001). However this is still unclear.

---

⋆ Based on data obtained at Las Campanas Observatory, Carnegie Institution.

One may expect that a detailed spectroscopic study of the nebulae might reveal chemical inhomogeneities due to the presence of material processed by the central stars, like in the case of the knots in A 30, A 78, and A 56. Such knots are H-deficient and consist mainly of He, C, O, and other heavy elements (Jacoby & Ford, 1983; Medina & Peña, 2000; Wesson et al., 2008). Those inhomogeneities would be the enriched knots proposed by Liu et al. as causing the large ADFs in PNe.

Ercolano et al. (2004) presented a complete study of the WRPNe NGC 1501, based on deep optical spectroscopy and detailed 3D photoionization modelling. They found a large ADF for this nebula and argued that the presence of a hydrogen-deficient metal-rich component is necessary to explain it. They also proposed that material ejected from the surface of the AGB precursor in a late thermal pulse event could be the possible origin of such knots.

The aim of the present study is to analyze deep high spectral resolution data of some WRPNe searching for evidence of chemical inhomogeneities through the determination of ADFs. Searching for abundance variations across the face of a nebula requires: good-quality spatially resolved spectra and a reliable procedure to correct for the unseen ions. For this purpose we have searched for WRPNe with available ORLs in a large sample of objects observed by M.P. at Las Campanas Observatory with the 6.4-m telescope Clay and the high-resolution spectrograph MIKE.

The original sample contains several WRPNe of all excitation classes, with central stars from early [WO] to late [WC] types. For this work, we have selected NGC 2867 and PB 8 because both nebulae present O II and C II ORLs strong enough to be analyzed; and additionally, PB 6, which present C II ORLs, but not O II ORLs. General characteristics of these objects are listed in Table 1. PB 6 and NGC 2867 were already studied with a similar purpose by Peña et al. (1998) using medium-resolution spectra. No abundance variations were found in their work.

In the following we present the results of our high-resolution data analysis. In § 2 we describe the observations and data reduction. In § 3 the observed and dereddened nebular line ratios are presented for the three objects in the different positions. In § 4 we analyze the data, deriving the physical conditions for all the available diagnostic ratios; and in § 5 we present the ionic abundances of visible ions from the CELs and ORLs and the total abundances of several elements. Finally, in § 6 we discuss our results, and in § 7 we present the summary and our conclusions.

## 2. Observations and data reduction

High spectral resolution data were obtained at Las Campanas Observatory (Carnegie Institution) with the 6.5-m telescope Clay and the double echelle spectrograph Magellan Inamori Kyocera Echelle (MIKE) on May 9 and 10, 2006. This spectrograph operates with two arms which allow to obtain a blue and a red spectrum simultaneously. We used the standard grating settings which provide a wavelength coverage from 3350 to 5050 Å in the blue and from 4950 to 9400 Å in the red. A complete description of MIKE performance can be found in Bernstein et al. (2003). We discarded spectral data from 9000–9400 Å due to abnormal flat-field structures that cannot be corrected.

The log of observations is presented in Table 2. The slit dimensions were 1″ wide (along the dispersion axis) and 5″ in the spatial direction. A binning of 2×2 was used. Thus the plate scale was 0.2608 arcsec/pix. Series of bias, as well as 'milky' flats and flats with the internal incandescent lamp were acquired for calibration. For wavelength calibrations a Th-Ar lamp was observed immediately after each observation. The spectral resolution varied in the blue from 0.14 to 0.17 Å ($\sim$10.8 km s$^{-1}$ in average) and in the red from 0.23 to 0.27 Å ($\sim$12.8 km s$^{-1}$ in average) as measured from the full width at half maximum (FWHM) of lines of the Th-Ar comparison lamp.

In Fig. 1 we present a portion of the echellogram around the H$\beta$ nebular line showing the spatially and spectroscopically resolved emission for PB 8, NGC 2867, and PB 6, as well as the extracted spectra. The limits of the extracted sections are indicated by black lines. We limited these sections to be 0.9″ wide in the spatial direction for two reasons: a) to avoid contamination of the intense stellar continua, and b) to fit the width of the flat-field exposures.

All the objects were observed at zenith distances smaller than or about 30$^o$ (as recommended by MIKE User Manual), covering an airmass range among 1.075 to 1.172, thus the atmospheric refraction was not expected to affect the spectra.

Observations of the standard HD 49798 were used for flux calibration. The standards HR 5501 and HR 4468 were also observed, but were discarded because their broad absorption Balmer lines make the continuum flux calibration very unreliable.

We performed standard reduction procedures on our data: 2D-echellograms were bias-subtracted and flat-fielded using IRAF[1] reduction packages. Spectra were extracted with extraction windows of 0.9″ (see above) and were flux calibrated. All the observed lines, not affected by the atmosphere, were measured.

## 3. Line Intensities and Reddening Correction

Line intensities were measured applying a single or a multiple Gaussian profile fit. In the case of some lines which present velocity structure, we integrated all the flux in the line between two given limits, over a local continuum estimated by eye. All these measurements were made with the SPLOT routine of the IRAF package. In some cases of very tight blends or blends with very bright telluric lines the analysis was performed via Gaussian fitting (or Voigt profiles in the case of sky emission lines) making use of the Starlink DIPSO software (Howarth & Murray, 1990). For each single or multiple Gaussian fit, DIPSO gives the fit parameter (radial velocity centroid, Gaussian sigma, FWHM, etc.) and their associated statistical errors.

All lines of a given spectrum were normalized to a particular bright emission line present in the common range of both, blue and red spectrum. In the cases of NGC 2867 and PB 8, we have used the He I $\lambda$5015 line. For PB 6, the reference line was [O III] $\lambda$5007. In order to produce a homogeneous set of line flux ratios, all of them were rescaled to the H$\beta$ flux. Some lines which were saturated in the long exposures were measured in the short ones and rescaled to the H$\beta$ flux.

Table 3 presents the emission line intensities measured in the three PNe. The first column presents the adopted laboratory wavelength, $\lambda_0$. The second and third columns present the ion and the multiplet number or series for each line. Columns 4, 7, 10, 13, and 16 present the heliocentric velocities of the different components measured. Dereddened intensity line ratios relative to H$\beta$, $I(\lambda)$, are presented in columns 5, 8, 11, 14, and 17. Finally, the observational errors (1$\sigma$) associated with the line fluxes with respect to H$\beta$ — in percentage — are also presented in columns 6, 9, 12, 15, and 18. These errors include the uncertainties in the flux measurement, flux calibration (estimated to be about 5%) and the error propagation in the reddening coefficient.

---

[1] IRAF is distributed by the National Optical Astronomy Observatories, which is operated by the Association of Universities for Research in Astronomy, Inc., under contract to the National Science Foundation.

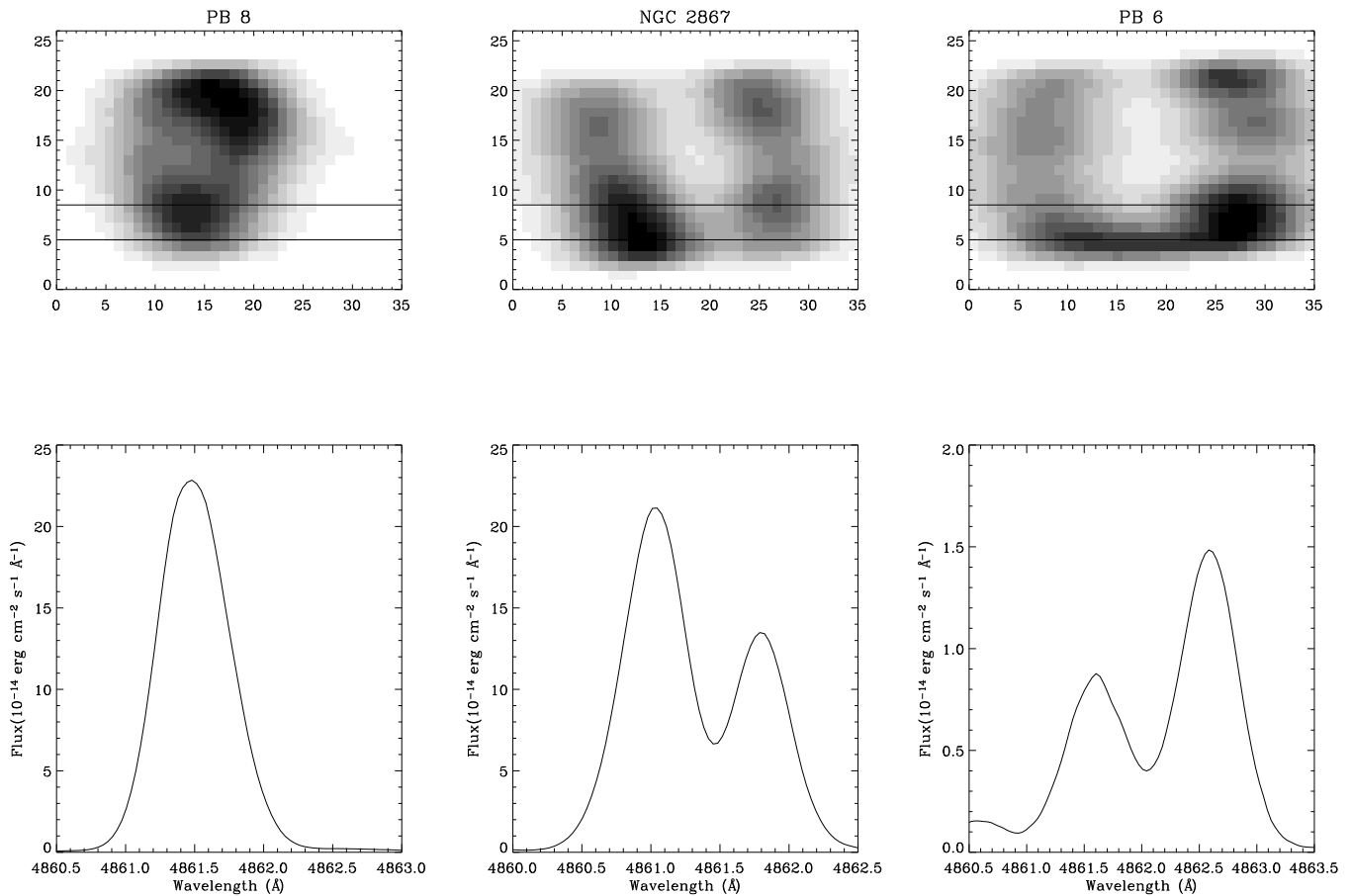

**Fig. 1.** Portion of the echellogram (up) and the extracted spectra (down) showing the spatially resolved H$\beta$ line for PB 8, NGC 2867, and PB 6. The limits of the extracted sections (0.9″ wide) are indicated by black lines.

A total of 211, 203, and 93 emission lines were measured in PB 8, NGC 2867, and PB 6, respectively. Most of the lines are permitted lines (see Table 3) of H I, He I, and He II, but also we have detected a considerable amount of heavy element permitted lines, such as O II, O III, N II, N III, C II, C III, C IV, and Ne II lines, whose analysis will be discussed in § 5.3. In several cases some identified lines were severely blended with telluric lines, making impossible their measurement. Other lines were strongly affected by atmospheric features in absorption or by internal reflections by charge transfer in the CCD, rendering their intensities unreliable.

The identification and adopted laboratory wavelengths of the lines were obtained following several previous identifications in the literature (see García-Rojas et al., 2004; Esteban et al., 2004; Zhang et al., 2005, and references therein).

We have assumed the standard dust extinction law for the Milky Way (R$_v$=3.1) parametrized by Seaton (1979) for our three objects. The reddening coefficients were derived by fitting the observed $I$(H Balmer lines)/$I$(H$\beta$) ratios — those not contaminated by telluric or other nebular emissions, or by absorption bands — and $I$(H Paschen lines)/$I$(H$\beta$), to the theoretical ones computed by Storey & Hummer (1995) for $T_e$ and $n_e$ previously estimated for the objects. In Table 4 we show the values obtained for the extinction coefficients c(H$\beta$) and the H I lines used for our three PNe and their kinematic components. There is an overall good agreement between our derived c(H$\beta$) values and those from previous determinations in these objects. Peña et al. (1998) found a c(H$\beta$) between 0.34 and 0.53 in three different aperture extractions in PB 6, and values between 0.35 and 0.37 for three aperture extractions for NGC 2867. For PB 8, our c(H$\beta$) is very consistent with that derived by Tylenda et al. (1992), c(H$\beta$)=0.38, however, Girard et al. (2007) found a relatively higher value of 0.68; this discrepancy could be due to difficulties found by these authors in separating the nebular from the stellar emission.

## 4. Physical Conditions

### 4.1. Temperatures and Densities

The large number of emission lines identified and measured in the spectra allows us to derive physical conditions using different emission line ratios. The temperatures and densities are presented in Table 5. Most of the determinations were carried out with the IRAF task TEMDEN of the package NEBULAR (Shaw & Dufour, 1995). We have updated the atomic data of our ver-sion of NEBULAR (2.12.2) to state-of-the-art transition probabilities and collisional strengths (see Table 6).

In the case of electron densities, ratios of CELs of several ions have been used. From inspection of Table 5, it seems that there are no apparent differences between densities for ions with low and high ionization potentials for PB 8 and PB 6, therefore, we have computed a weighted average value from the values obtained from [O II], [S II], [Cl III], and [Ar IV] ratios. For NGC 2867, it is apparent a slight gradient of the densities with the ionization potentials of the ions but, taking into account the uncertainties involved and the small dependency of our study with density stratification, we do not consider it important for our analysis. We have adopted a weighted average of $n_e$([O II]), $n_e$([S II]), $n_e$([Cl III]), and $n_e$([Ar IV]). We have excluded $n_e$(N I) in this object because this ion is representative of the very outer part of the nebulae, and probably does not coexist with most of the other ions.

The adopted density was used to derive $T_e$(N II), $T_e$(O II), $T_e$(S II), $T_e$(O III), and $T_e$(Ar III), and we iterated until convergence. Electron temperatures were derived from the ratio of CELs of several ions and making use of NEBULAR routines.

To obtain $T_e$(O II) it is necessary to subtract the contribution to [O II] $\lambda\lambda$7320+7330 due to recombination. Liu et al. (2000) found that this contribution can be fitted in the range $0.5 \leq T_e/10^4 \leq 1.0$ by:

$$\frac{I_R(7320+7330)}{I(H\beta)} = 9.36 \times (T_4)^{0.44} \times \frac{O^{++}}{H^+}, \quad (2)$$

where $T_4 = T/10^4$. Assuming the values derived for $O^{++}/H^+$ from ORLs and using this equation, we have estimated a contribution of approximately 59% of recombination to the observed line intensities in PB 8. This yields a $T_e$([O II])=7050 K, 4350 K lower than those derived without taking into account recombination contribution. In the two knots of PB 6 and NGC 2867, the different components of [O II] $\lambda\lambda$7320, 7330 lines are severely blended, making their individual intensities unreliable.

Liu et al. (2000) also determined that the contribution to the intensity of the $\lambda$5755 [N II] line due to recombination can be estimated from:

$$\frac{I_R(5755)}{I(H\beta)} = 3.19 \times (T_4)^{0.30} \times \frac{N^{++}}{H^+}, \quad (3)$$

in the range $0.5 \leq T_4 \leq 2.0$. Adopting the $N^{++}/H^+$ ratio obtained from recombination N II lines (see § 5.3), we have obtained a contribution of about 28% for PB 8. This yields a $T_e$([N II])=8900 K, 1200 K lower than that derived without recombination contribution, and which is 1850 K higher than the new computed $T_e$([O II]). In NGC 2867 and PB 6, we could not detect N II ORLs. From the semiforbidden N III] line measured by Peña et al. (1998) in the internal zones of NGC 2867 and PB 6, we have estimated $N^{++}/H^+$ of about 6×10$^{-5}$ and 1×10$^{-4}$ for NGC 2867 and PB 6, respectively. Assuming an ADF($N^{++}$) of about 2.5, we have estimated that the contributions to the intensity of [N II] $\lambda$5755 line are between 5-10 %. At the temperatures of our objects, these contributions will reduce our derived temperatures by a maximum value of 750 K. Taking into account that this value is only slightly larger than our adopted errors, and the uncertainty of the estimated $N^{++}/H^+$, we have decided not to apply the correction for NGC 2867 and PB 6 because it would not affect significantly our results.

We have computed $T_e$([S II]) for the two knots of NGC 2867. These temperature determinations are very uncertain because the [S II] $\lambda$4068.6 line is blended with C III $\lambda$4068.91, and [S II] $\lambda$4076.35 line is blended with O II multiplet 10 $\lambda$4075.86 line. We have adopted the theoretical relative strengths of the C III and O II lines to other observed lines of the same multiplet, in order to decontaminate [S II] lines. The values of $T_e$([S II]) we have obtained are about 3000 K lower than those obtained from N II lines, and taken into account their large uncertainties, we have not considered them in our analysis.

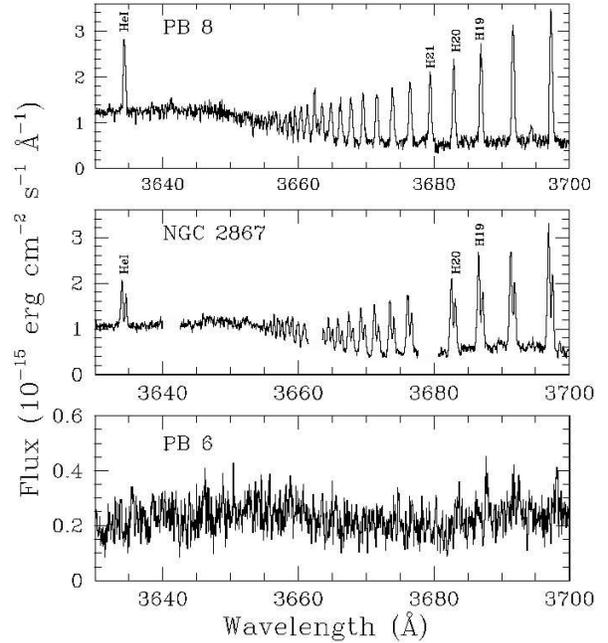

**Fig. 2.** Section of the echelle spectra showing the Balmer discontinuity in PB 8 (upper panel), NGC 2867 (medium panel) and PB 6 (lower panel). Gaps in NGC 2867 spectrum are due to the removal of features produced by charge transfer in the CCD due to the saturation of [O III] $\lambda\lambda$4959,5007 lines.

In NGC 2867 and PB 6, a large amount of oxygen is in the form of $O^{3+}$, so we have to take into account the contribution of recombination excitation to the auroral [O III] $\lambda$4363 line, which can be estimated from equation 3 of Liu et al. (2000):

$$\frac{I_R(4363)}{I(H\beta)} = 12.4 \times (T_4)^{0.79} \times \frac{O^{3+}}{H^+}. \quad (4)$$

We cannot compute directly the $O^{3+}$ from CELs or from ORLs (see § 5), but we can estimate it from helium ionic abundances using: $O^{3+}/H^+ = [(He/He^+)^{2/3} - 1] \times (O^+/H^+ + O^{++}/H^+)$ (Kingsburgh & Barlow, 1994), where $O^+/H^+$ and $O^{++}/H^+$ can be taken from Table 9. With this expression we have estimated a contribution due to recombination to the [O III] $\lambda$4363 line of ~2% for NGC 2867 and ~3% for PB 6, which have negligible effects in the determination of $T_e$([O III]).

Figure 2 shows the spectral regions near the Balmer limit in PB 8, NGC 2867 and PB 6. The discontinuities are clearly appreciated in PB 8 and NGC 2867. The Balmer discontinuity is defined as $I_c(Bac) = I_c(\lambda 3646^-) - I_c(\lambda 3646^+)$. The high spectral resolution of the spectra permits us to measure the continuum emission in zones very near the discontinuity, minimizing the possible contamination of other continuum contributions. The uncertainty in the derived continua is the standard mean deviation of the averaged continua. Finally, we have com-puted $T_e$(Bac) from the ratio of the Balmer Jump to H11 flux, using the relation proposed by Liu et al. (2001):

$$T_e(\text{Bac}) = 368 \times \left(1 + 0.259y^+ + 3.409y^{++}\right)\left(\frac{\text{BJ}}{\text{H11}}\right)^{-\frac{3}{2}} \quad (5)$$

where $y^+$ and $y^{++}$ correspond to the He$^+$/H$^+$ and He$^{++}$/H$^+$ ratios, respectively (see § 5.1).

Peimbert et al. (2002) developed a method to derive the helium temperature, $T_e$(He I), in the presence of temperature fluctuations (see § 5.1). Using their formulation, we have derived the $T_e$(He I) showed in Table 5 for PB 8, NGC 2867, and PB 6. The computed $T_e$(He I)'s are higher than those derived from H I.

We have assumed a 2-zone (for PB 8) and a 3-zone (for NGC 2867 and PB 6) ionization scheme for the calculation of ionic abundances (see § 5). The electron temperature obtained from [N II] lines was adopted as representative for the low ionisation zone designated as $T_e$(low). The average of electron temperatures obtained from [O III] and [Ar III] lines has been assumed as representative of the high ionization zone designated as $T_e$(high) (see Table 5).

A third ionization zone was considered in NGC 2867 and PB 6, due to the detection of the high ionization [Ne V] $\lambda$3425.87 line in both objects, which indicates the presence of a very high ionization zone in the inner parts of these PNe. We have computed $T_e$([Ar IV]) from the [Ar IV] $(\lambda 4711+\lambda 4740)/(\lambda 7170+\lambda 7263)$ $T_e$ sensitive line ratio. This ratio is valid in the range 2000 K<$T_e$<20000 K. Due to a strong blend between the [Ar IV] $\lambda$4740 line and a feature due to charge transfer in the CCD, the $T_e$ we have obtained from this ratio for component 1 of NGC 2867, is not reliable. For component 2, we only could determine an upper limit of about 20000 K. We have used the results of tailored photoionization models by Peña et al. (1998) to estimate $T_e$ in the inner zones of the nebula, which show an increase of about 2000 K between O$^{++}$ and O$^{3+}$ zones, so we have adopted a $T_e$ for the higher ionization zone of 13850±2000 K and 13600±2000 for components 1 and 2 of NGC 2867, respectively. For PB 6, the observed [Ar IV] line ratio was outside the validity limits of this diagnostic. Photoionization models of Peña et al. (1998) predicted an ionization structure in which O$^{3+}$ zone had a $T_e$ 4000 K larger than O$^{3+}$ zone; therefore, we have adopted a value of $T_e$(O$^{3+}$)=18000±2000 K for the two components of PB 6.

The effect of the presence of spatial temperature fluctuations will be discussed in § 6.1

## 5. Ionic and elemental abundances

### 5.1. He$^+$ and He$^{++}$ abundances

Several He I emission lines in the spectra of PB 8, and NGC 2867, and only 4 lines in the spectra of PB 6 were measured. These lines arise mainly from recombination but they can be affected by collisional excitation and self-absorption effects.

We have used the effective recombination coefficients of Storey & Hummer (1995) for H I and those computed by Porter et al. (2005), with the interpolation formulae provided by Porter et al. (2007) for He I. The collisional contribution was estimated from Sawey & Berrington (1993) and Kingdon & Ferland (1995), and the optical depth in the triplet lines were derived from the computations by Benjamin et al. (2002). We have determined the He$^+$/H$^+$ ratio from a maximum likelihood method (MLM, Peimbert et al., 2000, 2002). To determine $n_e$(He I), $T_e$(He I), He$^+$/H$^+$ ratio and the optical depth in the He I $\lambda$3889 line, ($\tau_{3889}$), self consistently, we have used the adopted density obtained from the CEL ratios for each object (see Table 5) and a set of $I$(He I)/$I$(H I) line ratios. For PB 8 and the two knots of NGC 2867, we have a total of 13 observational constraints (12 lines + $n_e$); each of these constraints depends upon the four unknown quantities, each dependence being unique. Finally, we have obtained the best value for the 4 unknowns and $t^2$ by minimizing $\chi^2$. The obtained $\chi^2$ parameters are showed in Table 7; these parameters indicate a reasonable goodness of the fits, taking into account the degrees of freedom in each case. The case of PB 6 is uncertain because we have not observed the He I $\lambda$3889 line, so the MLM cannot converge in order to reach a reasonable value of $\tau_{3889}$.

PB 8 is a relatively low ionization PNe, so there are no He II lines in its spectrum. On the other hand, we have measured several He II emission lines in both knots of NGC 2867 and PB 6. We have used the brightest lines to compute He$^{++}$/H$^+$ ratio, by using the recombination coefficients computed by Storey & Hummer (1995). There is a very good agreement between the results obtained from single lines, so we have finally adopted the He$^{++}$/H$^+$ average, weighted by the uncertainties of each individual line. Final results are presented in Table 7.

### 5.2. Ionic Abundances from CELs

Ionic abundances of several heavy metal ions were determined from CELs, using the IRAF package NEBULAR. $T_e$([N II]) was assumed for ions of lowest ionization potentials, N$^+$, O$^+$ and S$^+$; $T_e$([O III]) was assumed for ions with ionization potentials between 15 and 41 eV, O$^{++}$, Ne$^{++}$, S$^{++}$, Cl$^{++}$, Cl$^{3+}$, Ar$^{++}$ and Ar$^{3+}$; for the ions with higher ionization potentials, such as Ne$^{3+}$, Ne$^{4+}$ and Ar$^{4+}$, we adopted the temperature of the very high ionized zone estimated from photoionization models (see§ 4).

Ionic abundances are listed in Table 8 and correspond to the mean value of the abundances derived from all the individual lines of each ion observed (weighted by their relative strengths).

To derive the abundances for $t^2$>0.00 (see § 6.1) we used the abundances for $t^2$=0.00 and the formulation of by Peimbert (1967) and Peimbert & Costero (1969). For different $t^2$ values than assumed, it is possible to interpolate or extrapolate the values presented in Table 8.

### 5.3. Ionic Abundances from Recombination Lines

We have measured a large number of permitted lines of heavy element ions such as O II, O III, C II, C III, C IV, N II, N III, Si II and Ne II, many of them detected for the first time in these nebulae. Unfortunately, many permitted lines are affected by fluorescence effects or blended with telluric emission lines making their intensities unreliable. Detailed discussions on the mechanism of formation of the permitted lines until the twice ionized stage can be found in Esteban et al. (1998, 2004, and references therein). Discussion about mechanism of formation of N III, O III and C III lines can be found in Grandi (1976).

For the first time for these nebulae, we have been able to measure the ionic abundance ratios of several ions from pure recombination lines. We have computed the abundances using $T_e$(O III) and $n_e$ from Table 5. The atomic data (recombination coefficients) used for each ion are compiled in Table 6. In Tables 9 to 11 we present the abundances obtained for our objects from C II, C III, C IV, O II, N II and Ne II ORLs. Following Esteban et al. (1998) we have taken into account the abundances obtained from the intensity of each individual line and the abun-dances from the estimated total intensity of each multiplet (labelled as "Sum"); the "Sum" abundances were obtained by multiplying the sum of the intensities of the observed lines by the multiplet correction factor, which takes into account the unseen lines of the each multiplet by using their relative line strengths, $\log(gf)$. The $\log(gf)$'s have been constructed assuming that they are proportional to the population of their parent levels assuming LTE computation predictions from Wiese et al. (1996) for all the lines except for C IV for which we have adopted the $\log(gf)$'s given by the computations in the Atomic Line List v2.04[2]. For the adopted values, we have only considered the abundances marked as boldface in Tables 9 to 11.

Several permitted lines of C II have been measured in NGC 2867, and only one in PB 8 and PB 6. Lines of multiplets 6 and 17.04 are $3d-4f$ transitions and are, in principle, excited by pure recombination (see Grandi, 1976). In NGC 2867, values obtained from multiplet 6 and 17.04 are in very good agreement, within uncertainties, and shows clearly that the difference in the $C^{++}$ abundance betweeen the two components is real (see Table 10).

We have detected several lines of multiplets 1, 16 and 18 of C III in NGC 2867 and PB 6. The ground state of C III is a singlet ($2s^2\ ^1S$), so triplet lines, such multiplet 1 ($3s^3S-3p^3P$) and 16 ($4f^3F-5g^3G$) arise from pure recombination (Grandi, 1976). Also, the singlet C III $\lambda 4186.90$ line has a $^1G$ term as its upper level so, it is probably excited by recombination.

Two C IV lines were measured in the spectrum of NGC 2867 and one in the spectrum of PB 6. Grandi (1976) argued that both are produced mainly by recombination. The adopted values for $C^{4+}/H^+$ are listed in Tables 10 and 11.

We have measured a large number of O II lines in PB 8. Esteban et al. (1998, 2004) demonstrated that the best lines to compute the $O^{++}$ from O II lines are those of multiplets 1, 2, 10, 20 and $3d-4f$ transitions, because they are all excited by recombination, so we have only took into account abundances from these multiplets.

As it has been pointed out by Tsamis et al. (2003) and Ruiz et al. (2003), the upper levels of the transitions of multiplet 1 of O II are not in LTE for densities $n_e < 10000$ cm$^{-3}$, and the abundances derived from each individual line could differ by factors as large as 4. We have applied the NLTE corrections estimated by Peimbert et al. (2005) to our data and the abundances obtained from the individual lines are in good agreement and also agree with the abundance derived using the sum of all the lines of the multiplet, which is not affected by NLTE effects.

We have detected several lines of multiplets 2 and 5 of O III in the spectra of NGC 2867 and PB 6. Grandi (1976) showed that multiplet 2 lines are excited by Bowen fluorescence, so they are unreliable to derive the $O^{3+}/H^+$ ratio. Multiplet 5 of O III was not discussed by Grandi (1976), but the only line of this multiplet we have detected in NGC 2867, gives an $O^{3+}/H^+$ ratio very similar to that obtained from multiplet 2, so it is probably excited by other mechanism than recombination.

Several N II permitted lines corresponding to multiplets 3, 5, 19, 20, 24 and 28, have been measured in the spectrum of PB 8, but no one in the spectra of NGC 2867, nor in PB 6. Grandi (1976) and Escalante & Morisset (2005) discussed the formation mechanism of several N II permitted lines in the Orion nebula, and concluded that recombination cannot account for the observed intensities of most of them, being resonance fluorescence by line and starlight excitation the dominant mechanisms. Also, Liu et al. (2001) suggested that continuum fluorescence

[2] webpage at: http://www.pa.uky.edu/~peter/atomic/

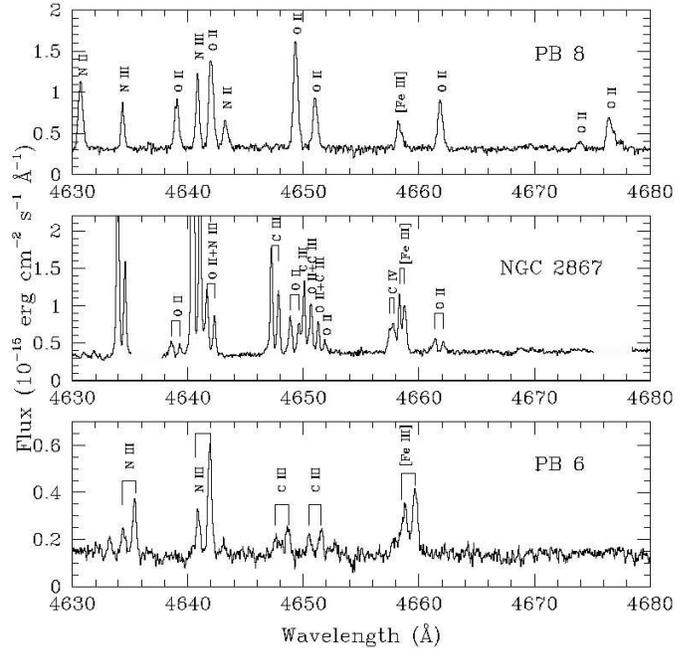

**Fig. 3.** Section of the echelle spectra showing the recombination emission lines of multiplet 1 of O II and multiplet 1 of C III. N II and N III emission lines in these plot are not excited by pure recombination (see text).

by starlight could be the excitation mechanism of N II permitted lines in PNe M 1−42 and M 2−36. On the other hand, the upper term of the N II $\lambda 4236.91$, $\lambda 4237.05$ and $\lambda 4241.78$ lines of multiplet 48 is $4f\ ^3F$, and cannot be populated by permitted resonance transitions, therefore this line should be excited mainly by recombination, so this is the value we adopt. The N$^{++}$ abundance obtained by co-adding the intensities of these lines, along with the characteristic wavelength of the whole multiplet, 4239.40 Å, are showed in Table 9.

We have measured several lines of multiplets 1 and 2 of N III in all our objects. Lines from both multiplets seem to be excited by the Bowen mechanism and are not reliable for abundance determinations (Grandi, 1976).

We have detected two $3d$-$4f$ transitions belonging to multiplet 57 of Ne II. These transitions are probably excited by recombination because they correspond to quartets and their ground level has doublet configuration (Esteban et al., 2004). For these transitions we have used effective recombination coefficients from recent calculations of Kisielius & Storey (unpublished), assuming LS-coupling. We have adopted the "sum" value derived from this multiplet: $12+\log(Ne^{++}/H^+)=8.28^{+0.12}_{-0.17}$ as representative of the Ne$^{++}$ abundance. This is the first time that Ne$^{++}/H^+$ has been derived from recombination lines for PB 8.

### 5.4. Total abundances

In order to correct for the unseen ionisation stages and then derive the total gaseous abundances of chemical elements in our PNe, we have adopted the set of ionisation correction factors (ICF) proposed by Kingsburgh & Barlow (1994); however, some special cases will be discussed. In Table 12 we show the total abundances obtained for PB 8 and NGC 2867 for $t^2=0.00$ and $t^2>0.00$. For PB 6, we have presented only abundances for $t^2$=0.00 because the derived $t^2$ values are very uncertain (see § 6.1.

### 5.4.1. Helium

The absence of He II lines in the spectra of PB 8 indicates that $He^{++}/H^+$ is negligible. However, the total helium abundance has to be corrected for the presence of neutral helium. Based on the ICF($He^0$) given by Peimbert et al. (1992) and with our data, the ICF($He^0$) amounts to 1.00, indicating that all the helium in this PN is once ionized.

For NGC 2867 and PB 6, the total helium abundance is the sum of the ionic abundances of $He^+$ and $He^{++}$.

### 5.4.2. Oxygen

We have derived the O/H ratio from both CELs and from ORLs. In the case of PB 8, no ICF was needed, because the absence of He II lines in the spectra of this PN, and the similarity between the ionization potentials of $He^+$ and $O^{++}$ is indicative of the absence of $O^{3+}$. We have, therefore, computed the total oxygen abundance in this nebula by adding the ionic abundance ratios of $O^+/H^+$ and $O^{++}/H^+$. As we have not detected O I ORLs, we have assumed that $O^+/H^+$ for $t^2>0.00$ is representative of the ORL abundance of this ion (see Table 12).

For NGC 2867 and PB 6, we applied the ICF scheme by Kingsburgh & Barlow (1994).

### 5.4.3. Carbon

For NGC 2867 and PB 6, we have computed ionic abundances of $C^{++}$, $C^{3+}$ and $C^{4+}$ from ORLs. It is generally assumed that $C^+/C = N^+/N$ so, to take into account for the unseen $C^+$ we have adopted the following ICF:

$$\frac{C}{H} = \left(\frac{1}{1+N^+/N}\right)\frac{C^{++}+C^{3+}+C^{4+}}{H^+}. \quad (6)$$

In the case of PB 8, we have only detected C II ORLs, so we have adopted the ICF proposed by Kingsburgh & Barlow (1994), which amounts to 1.02.

### 5.4.4. Other elements

We have adopted the usual ICF scheme for N, in which it is assumed that $N/O=N^+/O^+$ (see e.g. Kingsburgh & Barlow, 1994). For NGC 2867 and PB 6 this expression gives ICF's which are in excellent agreement with those computed from tailored photoionization models by Peña et al. (1998).

For neon, lines of three ionization stages ($Ne^{++}$, $Ne^{3+}$ and $Ne^{4+}$) were detected in NGC 2867 and PB 6. We have computed the total Ne abundance by adding the ionic abundances, assuming that $Ne^+/H^+$ is negligible, which is reasonable for high excitation PNe. For PB 8, we have only detected $Ne^{++}$ lines, so we have used the typical ICF proposed by Peimbert & Costero (1969):

$$\frac{Ne}{H} = \left(\frac{O^+ + O^{++}}{O^{++}}\right)\frac{Ne^{++}}{H^+}. \quad (7)$$

We measured lines of two ionisation stages of chlorine in NGC 2867: $Cl^{++}$ and $Cl^{3+}$. To take into account the $Cl^{4+}$ fraction, we have used the ICF proposed by Kwitter & Henry (2001), which is given by the expression:

$$\frac{Cl}{H} = \left(\frac{He}{He^+}\right)\frac{Cl^{++}+Cl^{3+}}{H^+}. \quad (8)$$

In PB 8, only lines of twice ionized chlorine, $Cl^{++}$ were detected. We have applied the formula obtained from photoionization models of Girard et al. (2007): ICF($Cl^{++}$)=(He/$He^+$)$^2$, to correct for the unseen ionization stages of chlorine; this formula, gives us an ICF=1.01, indicating that in this PN, almost all chlorine is twice ionized. In PB 6, we have not detected [Cl III] lines in the spectrum, but [Cl IV] lines were clearly detected. This indicates than in this object most of the $Cl^{++}$ has been ionized to $Cl^{3+}$. To correct for the unseen $Cl^{4+}$ we have applied the empirical relation by Kwitter & Henry (2001), obtaining values of the ICF of 6.84 for PB 6−1 and 5.06 for PB 6−2.

## 6. Discussion

### 6.1. Temperature variations and the abundance discrepancy factor

Torres-Peimbert et al. (1980) proposed the presence of spatial temperature fluctuations (parametrized by $t^2$) as the cause of the discrepancy between abundance calculations based on CELs and ORLs. This is due to the different dependence on the electron temperature of the CELs and ORLs emissivities. Assuming the validity of the temperature fluctuations paradigm, the comparison of the abundances determined from both kind of lines for a given ion should provide an estimation of $t^2$. Also, Peimbert (1971) proposed that there is a dichotomy between $T_e$ derived from the [O III] lines and from the hydrogen recombination continuum discontinuities, which is correlated with the discrepancy between CEL and ORL abundances (e.g. Peimbert & Costero, 1969; Torres-Peimbert et al., 1980; Liu et al., 2000; Tsamis et al., 2004), so the comparison between electron temperatures obtained from both methods is an additional indicator of $t^2$. A complete formulation of temperature fluctuations has been developed by Peimbert (1967), Peimbert & Costero (1969) and Peimbert (1971) (see also Peimbert et al., 2002; Ruiz et al., 2003). Several mechanisms have been proposed to explain the presence of temperature fluctuations in H II regions and/or PNe (see Esteban, 2002; Torres-Peimbert & Peimbert, 2003, and references therein) but to date, it is still unclear what mechanism could produce such temperature fluctuations.

On the other hand, from the ratios between ORL and CEL abundances, we obtained the ADF($O^{++}$), as defined in § 1, for PB 8 and NGC 2867. The values we have computed are: ADF($O^{++}$)=2.57±0.28 for PB 8, and ADF($O^{++}$)=1.49±0.16 and 1.77±0.22 for the two componentes of NGC 2867. These values are moderate and, in principle, could be due to the presence of spatial temperature fluctuations.

Assuming the validity of the temperature fluctuations paradigm and that this phenomenon produces the abundance discrepancy factor, we have estimated the values of the $t^2$ parameter from the ADFs obtained for $O^{++}$ in PB 8 and NGC 2867, and for $Ne^{++}$ for PB 8. In Table 13 we include the different $t^2$ values that produce the agreement between the abundance determinations obtained from CELs and ORLs of $O^{++}$ and $Ne^{++}$, as well as the values of $t^2$ obtained from the comparison between the value of $T_e$(O III) and the value of $T_e$(Bac). As it can be seen, the different $t^2$ values obtained are rather consistent within the errors. In Table 13, we also include the $t^2$ obtained from the application of a maximum likelihood method (MLM) to search for the physical conditions, including He$^+$/H$^+$ ratios and optical depths, that simultaneously fit all the measured lines of He I (see § 5.1). In the case of PB 6, the derived $t^2$ values are very uncertain because we could not measure enough He I lines in the spectrum of this PN in order to reach convergence of the MLM (see § 5.1). The final adopted values for $t^2$ parameter are showed in Table 13.

One fact apparently against the temperature fluctuations paradigm is the different $t^2$s we have found in the two components of NGC 2867, which translate into slight differences in the total abundances (see Table 12). Anyway, we have to take into account that we have adopted a $t^2$ parameter, which strongly favor the O$^{++}$ zone of this nebula, and does not necessarily represent the whole PN. This is probably due to we have sampling zones with different ionization degree in the analyzed knots. Unfortunately, the $t^2$ values that represent the whole nebula ($t^2$(He$^+$) and ($t^2$(Bac-CELs))) present large uncertainties, however, they are in pretty good agreement.

### 6.2. Comparison with other abundance determinations

Several works on the chemical content of these WRPNe have been found in the literature. In most of them the abundances were computed from the intensities of optical CELs, based in classical chemical analysis techniques (e.g. Kaler et al., 1991; Kingsburgh & Barlow, 1994; Peña et al., 1998, 2001; Girard et al., 2007), however, in other works, the abundances were also computed from tailored photoionization models (Henry et al., 1996; Peña et al., 1998).

For PB 8, we have computed O/H=5.8×10$^{-4}$ from CELs, which is intermediate between the values obtained by Kingsburgh & Barlow (1994) (O/H=4.9×10$^{-4}$) and Girard et al. (2007) (O/H=6.6×10$^{-4}$) but in relative good agreement, taking into account the uncertainties. Comparing the ratio of the other elements with respect to oxygen, the agreement with these authors is excellent, except in the case of S/O, for which we have found a ratio 3 times larger than that found by Girard et al. (2007). The O/H ratio derived from ORLs is much higher than solar (by almost a factor of 3). This is a somewhat puzzling result, because it is not expected O-enrichment in the evolution of the progenitor stars, descendants of low and intermediate mass stars (LIMS). Nevertheless, O-enrichment can not be absolutely ruled out; observational evidence has shown that PNe can be O-enriched (showing O/H larger than in H II regions) at least, at low metallicities (see Peña et al., 2007; Wang & Liu, 2008, and references therein).

In the case of NGC 2867, our derived O/H ratio from CELs (3.8×10$^{-4}$) is somewhat lower than previous estimates in the literature: O/H=6.0×10$^{-4}$, 4.4×10$^{-4}$ and 4.3×10$^{-4}$ by Kingsburgh & Barlow (1994), Girard et al. (2007) and Peña et al. (1998), respectively. Peña et al. (1998) pointed out that the high O/H ratio obtained by Kingsburgh & Barlow (1994), could be due to an overestimation in their [O III] λ5007/Hβ intensity ratio. Nevertheless, the relative abundances of N/O, Ne/O, S/O, Ar/O and Cl/O are in very good agreement, within the uncertainties, among the different authors.

There is a broad distribution in the O/H values reported in the literature for PB 6. Henry et al. (1996) obtained a value which is 0.25 dex higher than that obtained in this work. Peña et al. (1998) argued that this could be due to its adopted ICF scheme: (He$^+$+He$^{++}$)/He$^+$, that would lead to an overestimation of the He$^{++}$ zone contribution to the abundance of oxygen when the He$^{++}$ zone is large, as is the case of PB 6. The high N/O ratio of PB 6 has been confirmed by our data; our value of N/O=1.41 is in excellent agreement with previous determinations by Peña et al. (1998) (1.40) and Girard et al. (2007) (1.41). This high N/O ratio, in addition with the high He/H is a confirmation that original He and N have been enriched by nuclear reactions in the parental star of PB 6; therefore, it is a Type I PN. For the rest of elements, the agreement is poorer, but still consistent within the uncertainties.

The value of the abundances for these PNe are consistent with these objects being disk PNe.

### 6.3. The C/O ratio derived from ORLs and CELs

It is well known that there is a problem with the carbon abundance in PNe due to the large discrepancy found between abundances derived from the C II λ4267 ORL and the C III] λ1907+09 CEL (Rola & Stasińska, 1994; Peimbert et al., 1995). For this reason, the C/O derived for our objects deserves a separate mention. To our knowledge, this is the first time that C/O ratio is computed from pure recombination lines for PB 8 and NGC 2867. In particular, we have obtained C/O=0.49±0.05 for PB 8, and C/O=2.90±0.35 for NGC 2867.

The high C/O found for NGC 2867 is indicative of C-enrichment, which was also found by Peña et al. (1998). These authors, computed the C/O ratio for NGC 2867, using different combinations of UV carbon CELs, carbon ORLs, and optical and UV oxygen CELs, and pointed out that it was not possible to reconcile simultaneously the observed C/O ratios to the results obtained from tailored photoionization models. In particular, for NGC 2867, they found that the observed C III] λ1909/O III] λ1663 ratio gave a C/O ratio of 3.3±1.6, and the observed C III] λ1909/[O III] λ5007 ratio, a value of 3.1±1.3; both values being 3 to 7 times higher than those predicted by models. These results seem to discard uncertainties involved in the linking of UV and optical spectra, but still induce electron temperatures significantly lower than that deduced from photoionization models (Peña et al., 1998). In the following discussion we adopt (C/O)$_{CEL}$=of 3.2 as representative of this PN.

For PB 8, the only UV fluxes reported in the literature are [O III] λλ1661+67 and C III] λλ1907+09 lines, measured by Feibelman (2000); using these fluxes, we have obtained a C$^{++}$/O$^{++}$∼0.024, which is one order of magnitude smaller than that obtained from ORLs (C$^{++}$/O$^{++}$∼0.49±0.06). We have re-analyzed these spectra from the *IUE* archive database, and we have obtained a C$^{++}$/O$^{++}$∼0.17, still too low compared with the one computed from ORLs. Finally we have concluded that this discrepancy could be due to the faintness of [O III] λλ1661+67, which has been measured only 1-2σ over the signal to noise. By using UV fluxes of C III] and optical [O III] fluxes, we have obtained (C$^{++}$/O$^{++}$∼1.03 and (C/O)$_{CEL}$=1.00, which is the value we adopt as representative of this PNe.

In the case of PB 6, we could only compute the C/O ratio from the C II λ4267 ORL and the [O III] CEL, obtaining C/O∼9.1$^{+2.4}_{-1.9}$, which is consistent, within the errors, with the value of C/O=7.3±1.8 obtained by Peña et al. (1998) from the same lines, while from CELs they derived a C/O ratio between 2.4 and 2.6.

Interestingly, preliminary analysis of the chemistry of the stellar winds of the central stars performed with Potsdam NLTE expanding atmosphere models produced the following mass fractions: for PB 8, He:H:N:C:O are 81.5:15:1.5:1:1; for NGC 2867, He:C:O are 65:26:9, and for PB 6, He:C:O:N are 58.5:30:10:1.5 (Todt et al. in preparation).

In PB6 and NGC2867, the central star abundances correspond very well with the values calculated for other [WC] stars

(Koesterke, 2001) and they are in agreement with predictions from Very Late Thermal Pulse (VLTP) scenarios for these stars (Herwig, 2001). On the other hand the chemical composition of PB8 is very unusual showing an extremely low C abundance for a [WC] star. These results will be discussed in detail elsewhere (Todt et al. in preparation).

From the above results we find that the C/O ratios, by number, in the stellar wind are 3.9, 4.0 and 1.3 for NGC2867, PB6 and PB8 respectively. The two first values are quite large and similar to the large nebular values derived either from ORLs or from CELs for these nebulae (see Table 13). This indicates that the C produced by nucleosynthesis was already contaminating the AGB stellar atmosphere previously to the main body nebular ejection. There was no further C-enrichment in the star, relative to O, after the nebular material ejection, or both elements were enriched in lock step.

On the other hand, PB8 shows solar C/O (same C/O as the Orion nebula, Esteban et al., 2004), then it is evident that the nebula was ejected previously to the enrichment of the present atmospheric abundances. After the ejection, the star has continued producing C, but the present abundances have not reach the normal [WC] values. Evidently the evolution of this central star has followed a different path than other [WC]s.

Except for the works of Rola & Stasińska (1994) and Peimbert et al. (1995), there is a lack of systematic studies dedicated to explore the behavior of the C/O ratio derived from ORLs and CELs in PNe. In particular, there are only a few works devoted to study the role of ADFs and C/O ratios in WRPNe (Peña et al., 1998; Ercolano et al., 2004) and these roles are still unclear.

Ercolano et al. (2004) analyzed the WRPN NGC1501, finding a very large ADF($O^{++}$) (32) in this nebula. They argue that temperature fluctuations can not account for such large ADFs and propose that metal-rich knots could provide a solution to this problem. The central star of NGC1501 is a typical [WC 4] with atmospheric He:C:O abundances of 0.36:0.48:0.16 (mass fraction), probably produced in a VLTP scenario, similarly to the central stars of PB6 and NGC2867. However the nebula is not C-rich, having C/O=0.31. In this sense this object behaves similar to PB8 and different than PB6 and NGC2867. The observed nebula was ejected previously to the C-enrichment of the stellar surface. According to Ercolano et al. scenario for NGC1501, C-rich knots would have been ejected afterwards and would be partially mixed with the external nebula, producing a (C/O)$_{ORL}$ larger than (C/O)$_{CEL}$, however, (C/O)$_{CEL}$ for NGC 1501 has not been determined due to the extremely faint C III] line (Feibelman, 1998). Ercolano et al. (2004) suggest that a correlation between H-deficient stars and large ADFs in their nebulae would be expected.

In order to investigate this point, we have compiled some PN data from the literature (Liu et al., 2004; Tsamis et al., 2004; Wesson et al., 2005), discarding the PNe which do not belong to our Galaxy, and also NGC 40, which is strongly affected by aperture effects. These works are the largest compilations of PNe with C and O abundances derived from both ORLs and CELs. In figure 4 we show the ADF($O^{++}$) vs. $\log[(C/O)_{ORL}/(C/O)_{CEL}]$ for the objects in those samples; PNe with [WC] central star: NGC 5315 (square) and IC 2003 (diamond) are represented with bigger symbols; we have also included the data for our WRPNe: PB 8 and NGC 2867; it can be seen that most objects have (C/O)$_{ORL}$/(C/O)$_{CEL}$ > 1, except a few objects, being three of them PB 8, NGC 2867, and IC 2003, all with H-deficient central stars. Another behavior that is observed is that most of the objects with C/O$_{ORL}$ > 1 (filled symbols) have (C/O)$_{ORL}$/(C/O)$_{CEL}$ > 1, being NGC 2867 the only exception.

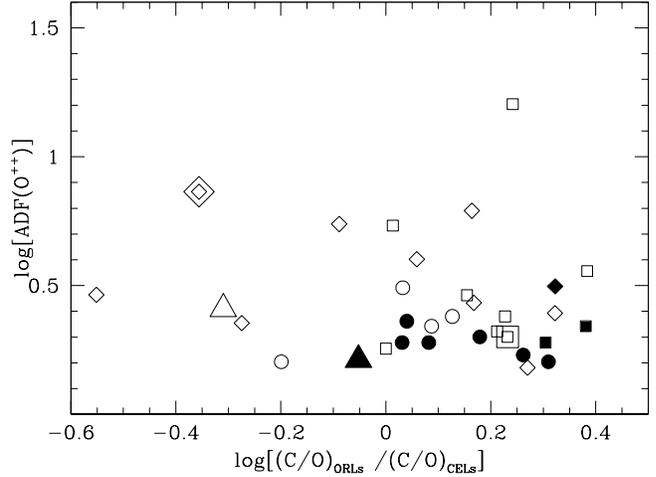

**Fig. 4.** Abundance discrepancy factor for ion $O^{++}$ vs. logarithm of the ratio of C/O obtained from ORLs and CELs. Circles are the objects from Liu et al. (2004); squares are from Tsamis et al. (2004) and diamonds are from Wesson et al. (2005). Filled symbols correspond to objects with high C/O ratio derived from ORLs (C/O >1.00). Triangles are our data for PB 8 and NGC 2867. Big symbols correspond to WRPNe data.

A puzzling observational fact is that altough theoretically one would expect VLTP scenario to produce C-rich material, studies in the suposed VLTP-produced knots in Abell 30 and Abell 58 by Wesson et al. (2003, 2008) showed C/O ratios less than 1. It could be that unexpected O-rich material might have been produced in the VLTP or that a different scenario than VLTP occured in Abell 30 and Abell 58 (Wesson et al., 2008). Similarly, we found C/O<1 for PB 8, but this object also shows (C/O)$_{ORL}$/(C/O)$_{CEL}$ < 1. The same last result has been found for NGC 2867 which added to our kinematical results (see § 6.4.2) in these PNe, seem to discard the C-rich knots ejected in a late thermal pulse event as the origin of the observed ADF in nebulae with H-deficient central stars.

A matter of concern is that in most of the objects, optical and UV observations do not cover identical volumes in the nebulae. Taking into account that C and O have been computed using a set of ionization correction factors, we have wondered if ionization structure effects would be affecting the derived C/O, but we found that C/O$_{ORL}$/(C/O)$_{CEL}$ are independent of the ionization degree. Another worry is that C/H derived from UV lines is highly uncertain due to extinction correction uncertainties. An underestimate of c(H$\beta$) would lead to a strong underestimate in the C abundance. A detailed treatment of error propagation would be of interest to evaluate the role of these uncertainties in the derived C/O ratio.

One last point we have to consider is the possible presence of spatial temperature fluctuations. Peimbert et al. (1995) analized the C and O abundances from the sample of PNe of Rola & Stasińska (1994) and argued that the differences between the measured $T_e(O^{++})$ and $T_e(C^{++})$ in a medium where $O^{++}$ and $C^{++}$ coexist could be due to the presence of spatial temperature fluctuations, and that in the presence of temperature inhomogeneities, the usual method of using $T_e$(5007/4363) would severely underestimate the $O^{++}$ and $C^{++}$ abundances as well as underestimate the $C^{++}/O^{++}$ ratio. Actually, any overestimation of the electron temperature would underestimate the $C^{++}$ abundance derived from the C III] $\lambda\lambda$1907+09 line, which emissivity is extremely sensitive to electron temperature.

It is clear for us, that a detailed and homogeneous study of the behavior of the C/O ratio derived both from ORLs and CELs, combined with an study of the central stars, in a large sample of WRPNe would shed some light on this issue.

### 6.4. Kinematic analysis

#### 6.4.1. Expansion velocities

Expansion velocities ($v_{exp}$) of the three nebulae have been measured as half the distance between maxima of double-peak profiles in the central region of [O III] $\lambda$5007 line (all the nebulae show double peak profiles). The values obtained are 38±4, 27±4 and 14±2 km s$^{-1}$ for PB 6, NGC 2867 and PB 8, respectively. These velocities show the same tendency as found for a large sample of WRPNe (Medina et al., 2006). That is, PB 6 and NGC 2867, which are ionized by early [WC] stars, show $v_{exp}$ much larger than the average found for PNe with no WR stars of similar stellar temperatures, while PB 8 shows a much lower $v_{exp}$, similar to the values of normal PNe with similar stellar temperature. The larger $v_{exp}$ in WRPNe is a consequence of the mechanical energy of the wind, blowing on the nebular shell.

#### 6.4.2. Heliocentric velocities of [O III] and O II lines

Peimbert et al. (2004) showed that the measurement of heliocentric velocities of [O III] and O II lines could provide evidence of the existence of H-poor clumps of high density and low temperature ejected at higher velocities and in a later stage than the main body of the nebula, as it has been proposed is happening in some PNe (see e.g. Guerrero & Manchado, 1996; Wesson et al., 2003). On the other hand, Tsamis et al. (2004), argued that the H-deficient knot model invoked to explain the ADFs in typical PNe, does not necessarily make specific predictions about kinematic patterns of the knots and the bulk of the nebula. Nevertheless, WRPNe are of particular interest because their central stars are hydrogen-deficient and, in principle, one could expect that late ejecta of material would be the origin of the proposed H-poor clumps. If this is the case, the velocities implied by [O III] and O II lines would be different.

We have found that the heliocentric velocities measured from [O III] and O II lines are very similar for PB 8 and NGC 2867, the objects in which both type of lines are measured. In PB 8, the average heliocentric velocity for the [O III] lines is 11.02±1.93 km s$^{-1}$ and for the O II lines is 13.28±0.97 km s$^{-1}$. For NGC 2867, the [O III] lines give −18.51±2.35 km s$^{-1}$ and 29.53±0.62 km s$^{-1}$, for blue and red components, respectively, and the O II lines give −16.76±0.67 km s$^{-1}$ and 31.45±0.97 km s$^{-1}$. The similarity obtained between the velocities argued against the scenario of H-poor clumps coming from a later ejecta at higher velocities than the main body of the PN. Nevertheless, a systematic study of the radial velocities of [O III] and O II lines would be desirable in order to lay down if phenomena occuring in WRPNe are related with the existence of H-poor clumps.

### 7. Summary

We present deep echelle spectroscopy in the 3250-9400 Å range of bright zones of [WC] PNe PB 8, NGC 2867 and PB 6. We have measured the intensity of an unprecedented number of lines per object. This is the most complete set of emission lines ever obtained for these three objects.

We have derived the physical conditions for each PN making use of several diagnostic line and continuum ratios. The chemical abundances have been derived using the intensity of collisionally excited lines (CELs) for a large number of ions of different elements. In addition, we have determined, for the first time in PB 8 and NGC 2867, the $O^{++}$ abundances from optical recombination lines (ORLs). Additionally, for NGC 2867 and PB 6, we derived the ORL abundances of $C^{++}$, $C^{3+}$ and $C^{4+}$. Finally, we also derived $N^{++}$ and $Ne^{++}$ abundance from ORLs for PB 8. From the comparison between CEL and ORL abundances, we have derived an ADF($O^{++}$)=2.57 for PB 8 and ADF($O^{++}$)=1.49 and 1.77 for the two components of NGC 2867. The derived ADF($O^{++}$) are in the range of typical ADFs observed in PNe, far from the large values found in other PN ionized by [WC] stars and "born again" PNe.

Assuming that the ADF and temperature fluctuations are related phenomena, we estimated the temperature fluctuations parameter, $t^2$ by comparing the $O^{++}$ (and in PB 8 also $Ne^{++}$) ionic abundances derived from ORLs to those derived from CELs. We have also obtained this parameter by applying a chi-squared method which minimizes the dispersion of $He^+/H^+$ ratios from individual lines and by comparing the electron temperatures derived from CELs to those derived from Balmer continua. All methods provided $t^2$ values consistent within the uncertainties. The adopted average value of $t^2$ has been used to correct the ionic abundances derived from CELs.

For PB 8 and NGC 2867, we have computed two sets of total abundances; the first one under the assumptions of homogeneous temperature and that CELs represent adequately the ionic abundances; and the second one assuming the presence of spatial temperature variations. For both sets we used a standard ICF scheme.

The inspection of C/O ratios in our WRPNe reveals that NGC 2867 has a high C/O ratio, derived from both, ORLs and CELs. Also, PB 6 has a large C/O derived from CELs intensities in the literature. On the contrary, PB 8 has a C/O ratio less than 1, revealing different chemical enrichment paths for the three objects. We also obtain that $(C/O)_{ORL}/(C/O)_{CEL} < 1$ in PB 8 and NGC 2867, and to that heliocentric velocities of the O II and [O III] lines of our three PNe are very similar, seem to discard the "C-rich knots ejected in a late thermal pulse" scenario as the origin of the observed ADFs. An alternative explanation could be the presence of spatial temperature fluctuations, that would lead the $C^{++}$ abundance from collisionally excited UV emission lines to be underestimated. A systematic study of C/O ratio, from both ORLs and CELs in WRPNe should be developed in order to clarify if H-deficient stars have any role in the ADF problem.

The expansion velocities of PB 8, NGC 2867 and PB 6 are consistent with the general behavior for WRPNe, which in general have $v_{exp}$ larger than the average found for PNe with similar stellar temperatures, due to the presence of strong stellar winds.

*Acknowledgements.* J. G.-R. is supported by a UNAM postdoctoral grant. M. Peña is grateful to DAS, Universidad de Chile, for hospitality during a sabbatical stay when part of this work was performed. M. P. gratefully acknowledges financial support from FONDAP-Chile and DGAPA-UNAM. We thank the referee for his/her comments and suggestions. The authors want to thank M. Peimbert for several discussions and suggestions and to H. Todt for providing results on the analysis of central stars of WRPNe previous to publication. This work received partial support from DGAPA/UNAM (grants IN118405, IN112708, 115807) and Conacyt-Mexico (grant #43121).

**Table 1.** General properties of the studied objects and sources of the UV data.

| Object (PN G) | Central star[a,b] | | | Diam.[c] | Log total |
|---|---|---|---|---|---|
| | $m_V$ | Spec. type | $T_*$(kK) | (″) | $I(H\beta)^{(2)}$ |
| PB 8 (292.4+04.1) | 13.8 | [WC 5-6] | 58 | 5 | −11.41 |
| NGC 2867 (278.1−05.9) | 15.7 | [WO 2] | 155 | 14 | −10.58 |
| PB 6 (278.8+4.9) | 17.6 | [WO 1] | 158 | 11 | −11.87 |

[a] Stellar magnitude and stellar type temperature as given by Acker & Neiner (2003)
[b] Temperature from Todt et al. (in preparation)
[c] Nebular diameter and total H$\beta$ flux from Acker et al. (1992)

**Table 2.** Log of observations.

| Object (PN G) | Obs. date | Exp. time (s) |
|---|---|---|
| PB 8 (292.4+04.1) | May 9, 2006 | 5, 900, 600, 300 |
| NGC 2867 (278.1−05.9) | May 9, 2006 | 5, 900, 900 |
| PB 6 (278.8+4.9) | May 10, 2006 | 600, 900 |

**Table 3.** Reddening corrected line ratios (F(Hβ) = 100) and line identifications.

| λ₀ (Å) | Ion | Mult. | PB8 | | | NGC 2867–1 | | | NGC 2867–2 | | | PB6–1 | | | PB6–2 | | |
|---|---|---|---|---|---|---|---|---|---|---|---|---|---|---|---|---|---|
| | | | $V_{rad}$ | $I(\lambda)$ [a] | Err(%) | $V_{rad}$ | $I(\lambda)$ [b] | Err(%) | $V_{rad}$ | $I(\lambda)$ [b] | Err(%) | $V_{rad}$ | $I(\lambda)$ [c] | Err(%) | $V_{rad}$ | $I(\lambda)$ [c] | Err(%) |
| 3425.87 | [Ne v] | | ... | ... | ... | -10 | 0.547 | 7 | 23 | 0.434 | 11 | 24 | 133.561 | 6 | 86 | 146.994 | 6 |
| 3428.65 | O III | 15 | ... | ... | ... | -14 | 2.706 | 6 | 26 | 2.575 | 6 | ... | ... | ... | ... | ... | ... |
| 3444.07 | O III | 15 | ... | ... | ... | -13 | 17.606 | 6 | 28 | 15.974 | 6 | 20 | 11.502 | 9 | 84 | 14.875 | 7 |
| 3447.59 | He I | 7 | 15 | 0.506 | 9 | ... | ... | ... | ... | ... | ... | ... | ... | ... | ... | ... | ... |
| 3487.73 | He I | 42 | 9 | 0.132 | 21 | ... | ... | ... | ... | ... | ... | ... | ... | ... | ... | ... | ... |
| 3530.50 | He I | 36 | 11 | 0.252 | 13 | ... | ... | ... | ... | ... | ... | ... | ... | ... | ... | ... | ... |
| 3554.42 | He I | 34 | 12 | 0.241 | 13 | -24 | 0.201 | 12 | 33 | 0.215 | 17 | ... | ... | ... | ... | ... | ... |
| 3587.28 | He I | 32 | 12 | 0.404 | 10 | ... | ... | ... | ... | ... | ... | ... | ... | ... | ... | ... | ... |
| 3613.64 | He I | 13 | 12 | 0.655 | 8 | -23 | 0.263 | 10 | 32 | 0.324 | 13 | ... | ... | ... | ... | ... | ... |
| 3634.25 | He I | 28 | 11 | 0.575 | 8 | -21 | 0.395 | 8 | 31 | 0.399 | 11 | ... | ... | ... | ... | ... | ... |
| 3660.28 | H I | H32 | 11 | 0.348 | 10 | ... | ... | ... | ... | ... | ... | ... | ... | ... | ... | ... | ... |
| 3661.22 | H I | H31 | 9 | 0.346 | 10 | ... | ... | ... | ... | ... | ... | ... | ... | ... | ... | ... | ... |
| 3662.26 | H I | H30 | 13 | 0.540 | 8 | ... | ... | ... | ... | ... | ... | ... | ... | ... | ... | ... | ... |
| 3663.40 | H I | H29 | 7 | 0.480 | 9 | ... | ... | ... | ... | ... | ... | ... | ... | ... | ... | ... | ... |
| 3664.68 | H I | H28 | 8 | 0.444 | 9 | ... | ... | ... | ... | ... | ... | ... | ... | ... | ... | ... | ... |
| 3666.10 | H I | H27 | 11 | 0.418 | 9 | ... | ... | ... | ... | ... | ... | ... | ... | ... | ... | ... | ... |
| 3667.68 | H I | H26 | 9 | 0.438 | 9 | -20 | 0.487 | 7 | 32 | 0.528 | 9 | ... | ... | ... | ... | ... | ... |
| 3669.47 | H I | H25 | 10 | 0.454 | 9 | -20 | 0.473 | 7 | 28 | 0.554 | 9 | ... | ... | ... | ... | ... | ... |
| 3671.48 | H I | H24 | 9 | 0.502 | 9 | -20 | 0.517 | 7 | 31 | 0.600 | 9 | ... | ... | ... | ... | ... | ... |
| 3673.76 | H I | H23 | 11 | 0.532 | 8 | -20 | 0.575 | 7 | 29 | 0.616 | 9 | ... | ... | ... | ... | ... | ... |
| 3676.37 | H I | H22 | 10 | 0.600 | 8 | -20 | 0.672 | 6 | 30 | 0.665 | 8 | ... | ... | ... | ... | ... | ... |
| 3679.36 | H I | H21 | 9 | 0.659 | 8 | ... | ... | ... | ... | ... | ... | ... | ... | ... | ... | ... | ... |
| 3682.81 | H I | H20 | 11 | 0.796 | 7 | -20 | 0.807 | 6 | 30 | 0.938 | 7 | ... | ... | ... | ... | ... | ... |
| 3686.83 | H I | H19 | 10 | 0.866 | 7 | -20 | 1.070 | 6 | 32 | 1.177 | 7 | ... | ... | ... | ... | ... | ... |
| 3691.56 | H I | H18 | 10 | 1.035 | 7 | -20 | 1.153 | 6 | 30 | 1.202 | 7 | ... | ... | ... | ... | ... | ... |
| 3694.22 | Ne II | 1 | 10 | 0.104 | 24 | ... | ... | ... | ... | ... | ... | ... | ... | ... | ... | ... | ... |
| 3697.15 | H I | H17 | 11 | 1.188 | 7 | -19 | 1.284 | 6 | 30 | 1.266 | 7 | ... | ... | ... | ... | ... | ... |
| 3703.86 | H I | H16 | 10 | 1.371 | 6 | -19 | 1.344 | 6 | 28 | 1.246 | 7 | ... | ... | ... | ... | ... | ... |
| 3705.04 | He I | 25 | 8 | 0.912 | 7 | -25 | 0.626 | 7 | 29 | 0.564 | 9 | ... | ... | ... | ... | ... | ... |
| 3707.25 | O III | 14 | ... | ... | ... | -15 | 0.055 | 30 | 23 | 0.056 | : | ... | ... | ... | ... | ... | ... |
| 3711.97 | H I | H15 | 11 | 1.577 | 6 | -19 | 1.511 | 6 | 30 | 1.659 | 6 | 19 | 1.894 | 28 | 80 | 2.036 | 19 |
| 3712.74 | O II | 3 | 31 | 0.198 | 15 | ... | ... | ... | ... | ... | ... | ... | ... | ... | ... | ... | ... |
| 3715.08 | O III | 14 | ... | ... | ... | -12 | 0.126 | 16 | 28 | 0.152 | 22 | ... | ... | ... | ... | ... | ... |
| 3721.83 | [S III] | 2F | 16 | 2.383 | 6 | -23 | 2.821 | 5 | 23 | 3.707 | 6 | 9 | 2.365 | 24 | 76 | 4.534 | 11 |
| 3721.93 | H I | H14 | * | * | * | * | * | * | * | * | * | * | * | * | * | * | * |
| 3726.03 | [O II] | 1F | 20 | 17.103 | 6 | -26 | 47.486 | 5 | 35 | 70.504 | 6 | 14 | 26.402 | 6 | 88 | 46.332 | 5 |
| 3728.82 | [O II] | 1F | 18 | 9.450 | 6 | -33 | 25.663 | 5 | 31 | 41.602 | 6 | 12 | 16.366 | 7 | 86 | 28.055 | 6 |
| 3734.37 | H I | H13 | 10 | 2.469 | 6 | -20 | 2.300 | 5 | 30 | 2.447 | 6 | 17 | 1.705 | 31 | 79 | 2.559 | 16 |
| 3749.48 | O II | 3 | 10 | 0.281 | 11 | ... | ... | ... | ... | ... | ... | ... | ... | ... | ... | ... | ... |
| 3750.15 | H I | H12 | 11 | 2.872 | 6 | -18 | 2.945 | 5 | 31 | 2.767 | 6 | 14 | 1.382 | 37 | 74 | 0.803 | : |
| 3754.69 | O III | 2 | ... | ... | ... | -14 | 0.612 | 7 | 27 | 0.732 | 8 | 13 | 0.844 | : | 81 | 2.243 | 18 |
| 3759.87 | O III | 2 | ... | ... | ... | -13 | 2.899 | 5 | 26 | 2.648 | 6 | 18 | 3.035 | 19 | 82 | 3.429 | 13 |
| 3770.63 | H I | H11 | 11 | 3.609 | 6 | -19 | 4.000 | 5 | 29 | 4.094 | 6 | 19 | 3.584 | 17 | 81 | 3.858 | 12 |
| 3774.02 | O III | 2 | ... | ... | ... | -13 | 0.231 | 10 | 25 | 0.298 | 13 | ... | ... | ... | ... | ... | ... |
| 3781.72 | He II | 4.21 | ... | ... | ... | -15 | 0.105 | 18 | 24 | 0.100 | 31 | ... | ... | ... | ... | ... | ... |
| 3791.27 | O III | 2 | ... | ... | ... | -13 | 0.224 | 10 | 28 | 0.265 | 14 | ... | ... | ... | ... | ... | ... |

| | | | PB8 | | | NGC 2867–1 | | | NGC 2867–2 | | | PB6–1 | | | PB6–2 | | |
|---|---|---|---|---|---|---|---|---|---|---|---|---|---|---|---|---|---|
| $\lambda_0$ (Å) | Ion | Mult. | $V_{rad}$ | $I(\lambda)$ [a] | Err(%) | $V_{rad}$ | $I(\lambda)$ [b] | Err(%) | $V_{rad}$ | $I(\lambda)$ [b] | Err(%) | $V_{rad}$ | $I(\lambda)$ [c] | Err(%) | $V_{rad}$ | $I(\lambda)$ [c] | Err(%) |
| 3797.63 | [S III] | 2F | 32 | 4.748 | 6 | 3 | 5.338 | 5 | 51 | 5.403 | 6 | 39 | 4.490 | 14 | 100 | 4.939 | 10 |
| 3797.90 | H I | H10 | * | * | * | * | * | * | * | * | * | * | * | * | * | * | * |
| 3813.50 | He II | 4.19 | ... | ... | ... | -11 | 0.149 | 13 | 28 | 0.211 | 17 | ... | ... | ... | ... | ... | ... |
| 3819.61 | He I | 22 | 12 | 1.498 | 6 | -20 | 1.000 | 6 | 34 | 1.688[e] | 6 | ... | ... | ... | ... | ... | ... |
| 3831.66 | S II | | 7 | 0.045 | : | ... | ... | ... | ... | ... | ... | ... | ... | ... | ... | ... | ... |
| 3833.57 | He I | 62 | 7 | 0.113 | 22 | 5 | 0.170 | 12 | 43 | 0.203 | 17 | ... | ... | ... | ... | ... | ... |
| 3835.39 | H I | H9 | 11 | 6.784 | 6 | -19 | 7.220 | 5 | 30 | 7.506 | 5 | 18 | 8.195 | 9 | 80 | 7.778 | 8 |
| 3837.91 | S III | 5 | ... | ... | ... | -52 | 1.651 | 5 | -2 | 1.603 | 6 | ... | ... | ... | ... | ... | ... |
| 3838.09 | He I | 61 | 31 | 0.149 | 18 | ... | ... | ... | ... | ... | ... | ... | ... | ... | ... | ... | ... |
| 3856.02 | Si II | 1 | 16 | 0.175 | 15 | ... | ... | ... | ... | ... | ... | ... | ... | ... | ... | ... | ... |
| 3856.13 | O II | 12 | * | * | * | ... | ... | ... | ... | ... | ... | ... | ... | ... | ... | ... | ... |
| 3858.07 | He II | 4.17 | ... | ... | ... | -12 | 0.201 | 11 | 26 | 0.228 | 16 | ... | ... | ... | ... | ... | ... |
| 3862.59 | Si II | 1 | 9 | 0.128 | 19 | ... | ... | ... | ... | ... | ... | ... | ... | ... | ... | ... | ... |
| 3867.49 | He I | 20 | 15 | 0.149 | 17 | ... | ... | ... | ... | ... | ... | ... | ... | ... | ... | ... | ... |
| 3868.75 | [Ne III] | 1F | 12 | 19.164 | 6 | -18 | 102.625 | 5 | 31 | 109.206 | 5 | 15 | 97.317 | 6 | 84 | 132.395 | 5 |
| 3871.82 | He I | 60 | 8 | 0.122 | 20 | -25 | 0.093 | 19 | 27 | 0.084 | 35 | ... | ... | ... | ... | ... | ... |
| 3888.65 | He I | 2 | 29 | 19.892 | 6 | -19 | 11.195[d] | 5 | 23 | 26.540[d] | 5 | ... | ... | ... | ... | ... | ... |
| 3889.05 | H I | H8 | * | * | * | -8 | 14.729[d] | 5 | 32 | 9.918[d] | 5 | ... | ... | ... | ... | ... | ... |
| 3918.98 | C II | 4 | 11 | 0.162 | 16 | ... | ... | ... | ... | ... | ... | ... | ... | ... | ... | ... | ... |
| 3920.68 | C II | 4 | 11 | 0.269 | 11 | ... | ... | ... | ... | ... | ... | ... | ... | ... | ... | ... | ... |
| 3923.48 | He II | 4.15 | ... | ... | ... | -11 | 0.278 | 9 | 27 | 0.332 | 12 | 20 | 1.870 | 28 | 83 | 1.218 | 28 |
| 3926.53 | He I | 58 | 15 | 0.213 | 13 | ... | ... | ... | ... | ... | ... | ... | ... | ... | ... | ... | ... |
| 3964.73 | He I | 5 | 13 | 1.419 | 6 | -20 | 0.556 | 7 | 33 | 1.057[e] | 7 | ... | ... | ... | ... | ... | ... |
| 3967.46 | [Ne III] | 1F | 13 | 5.689 | 6 | -17 | 30.801 | 5 | 31 | 32.816 | 5 | 16 | 30.148 | 6 | 88 | 21.150 | 6 |
| 3970.07 | H I | H7 | 12 | 14.466 | 6 | -17 | 15.708 | 5 | 31 | 15.988 | 5 | 20 | 17.500 | 7 | 82 | 16.171 | 6 |
| 4009.26 | He I | 55 | 12 | 0.260 | 12 | -21 | 0.213 | 10 | 35 | 0.467 | 9 | ... | ... | ... | ... | ... | ... |
| 4025.60 | He II | 4.13 | ... | ... | ... | -9 | 0.541 | 7 | 26 | 3.665 | 5 | 24 | 2.377 | 22 | 77 | 2.521 | 15 |
| 4026.21 | He I | 18 | 12 | 3.116 | 6 | -19 | 2.040 | 5 | 32 | 2.114 | 6 | 32 | 3.996 | 15 | 82 | 1.091 | 30 |
| 4041.31 | N II | 39 | 16 | 0.065 | 32 | ... | ... | ... | ... | ... | ... | ... | ... | ... | ... | ... | ... |
| 4060.22 | [F IV] | | ... | ... | ... | -12 | 0.085 | 20 | 26 | 0.089 | 32 | ... | ... | ... | ... | ... | ... |
| 4067.94 | C III | 16 | ... | ... | ... | -14 | 0.151 | 13 | 24 | 0.328 | 12 | ... | ... | ... | ... | ... | ... |
| 4068.60 | [S II] | 1F | 7 | 0.223 | : | 9 | 1.580 | 6 | 35 | 2.316 | 6 | ... | ... | ... | ... | ... | ... |
| 4068.91 | C III | 16 | ... | ... | ... | * | * | * | * | * | * | 2 | 2.374 | 22 | 63 | 2.836 | 14 |
| 4069.62 | O II | 10 | 25 | 0.391 | 9 | ... | ... | ... | ... | ... | ... | ... | ... | ... | ... | ... | ... |
| 4069.89 | O II | 10 | * | * | * | ... | ... | ... | ... | ... | ... | ... | ... | ... | ... | ... | ... |
| 4070.26 | C III | 16 | ... | ... | ... | -13 | 0.407 | 7 | 26 | 0.386 | 10 | ... | ... | ... | ... | ... | ... |
| 4072.15 | O II | 10 | 13 | 0.265 | 11 | -15 | 0.122 | 15 | 32 | 0.171 | 19 | ... | ... | ... | ... | ... | ... |
| 4075.86 | O II | 10 | 15 | 0.275 | 11 | 10 | 0.619 | 6 | 71 | 0.641 | 8 | ... | ... | ... | ... | ... | ... |
| 4076.35 | [S II] | 1F | ... | ... | ... | * | * | * | * | * | * | ... | ... | ... | ... | ... | ... |
| 4085.11 | O II | 10 | 10 | 0.086 | 26 | ... | ... | ... | ... | ... | ... | ... | ... | ... | ... | ... | ... |
| 4089.29 | O II | 48 | 13 | 0.102 | 22 | ... | ... | ... | ... | ... | ... | ... | ... | ... | ... | ... | ... |
| 4092.93 | O II | 10 | 15 | 0.038 | : | ... | ... | ... | ... | ... | ... | ... | ... | ... | ... | ... | ... |
| 4097.22 | O II | 20 | 20 | 0.281 | 11 | -6 | 0.976 | 6 | 36 | 0.753 | 7 | ... | ... | ... | ... | ... | ... |
| 4097.26 | O II | 48 | * | * | * | * | * | * | * | * | * | ... | ... | ... | ... | ... | ... |
| 4097.33 | N III | 1 | ... | ... | ... | * | * | * | * | * | * | 17 | 2.022 | 25 | 83 | 2.092 | 17 |
| 4100.04 | He II | 4.12 | ... | ... | ... | -10 | 0.501 | 7 | 29 | 0.653 | 8 | 21 | 2.492 | 21 | 81 | 2.121 | 17 |
| 4101.74 | H I | H6 | 12 | 24.285 | 5 | -17 | 24.747 | 5 | 31 | 26.438 | 5 | 20 | 25.499 | 6 | 81 | 25.665 | 6 |

**Table 3.** continued.

| | | | PB8 | | | NGC 2867–1 | | | NGC 2867–2 | | | PB6–1 | | | PB6–2 | | |
|---|---|---|---|---|---|---|---|---|---|---|---|---|---|---|---|---|---|
| $\lambda_0$ (Å) | Ion | Mult. | $V_{rad}$ | $I(\lambda)$ [a] | Err(%) | $V_{rad}$ | $I(\lambda)$ [b] | Err(%) | $V_{rad}$ | $I(\lambda)$ [b] | Err(%) | $V_{rad}$ | $I(\lambda)$ [c] | Err(%) | $V_{rad}$ | $I(\lambda)$ [c] | Err(%) |
| 4103.43 | N III | 1 | ... | ... | ... | -16 | 0.980 | 6 | 26 | 0.264 | 13 | 5 | 2.637[h] | : | 78 | 1.408 | 24 |
| 4110.79 | O II | 20 | 17 | 0.147 | 17 | ... | ... | ... | ... | ... | ... | ... | ... | ... | ... | ... | ... |
| 4119.22 | O II | 20 | 14 | 0.087 | 25 | ... | ... | ... | ... | ... | ... | ... | ... | ... | ... | ... | ... |
| 4120.82 | He I | 16 | 8 | 0.274 | 11 | -17 | 0.171 | 12 | 33 | 0.172 | 18 | ... | ... | ... | ... | ... | ... |
| 4121.46 | O II | 19 | 15 | 0.163 | 16 | ... | ... | ... | ... | ... | ... | ... | ... | ... | ... | ... | ... |
| 4132.80 | O II | 19 | 14 | 0.202 | 13 | ... | ... | ... | ... | ... | ... | ... | ... | ... | ... | ... | ... |
| 4143.76 | He I | 53 | 13 | 0.430 | 8 | -20 | 0.261 | 9 | 32 | 0.323 | 12 | 14 | 0.430 | : | 68 | 1.211 | 27 |
| 4145.90 | O II | 106 | 27 | 0.039 | : | ... | ... | ... | ... | ... | ... | ... | ... | ... | ... | ... | ... |
| 4146.08 | O II | 106 | * | * | * | ... | ... | ... | ... | ... | ... | ... | ... | ... | ... | ... | ... |
| 4153.30 | O II | 19 | 14 | 0.250 | 12 | ... | ... | ... | ... | ... | ... | ... | ... | ... | ... | ... | ... |
| 4156.53 | O II | 19 | 9 | 0.106 | 22 | ... | ... | ... | ... | ... | ... | ... | ... | ... | ... | ... | ... |
| 4168.97 | He I | 52 | 27 | 0.161 | 16 | ... | ... | ... | ... | ... | ... | ... | ... | ... | ... | ... | ... |
| 4169.22 | O II | 36 | * | * | * | ... | ... | ... | ... | ... | ... | ... | ... | ... | ... | ... | ... |
| 4185.45 | O II | 36 | 14 | 0.065 | 32 | ... | ... | ... | ... | ... | ... | ... | ... | ... | ... | ... | ... |
| 4186.90 | C III | 18 | ... | ... | ... | -12 | 0.227 | 10 | 27 | 0.239 | 14 | 17 | 0.807 | : | 84 | 0.948 | 32 |
| 4189.58 | O II | 36 | 13 | 0.077 | 28 | ... | ... | ... | ... | ... | ... | ... | ... | ... | ... | ... | ... |
| 4189.79 | O II | 36 | * | * | * | ... | ... | ... | ... | ... | ... | ... | ... | ... | ... | ... | ... |
| 4199.83 | He II | 4.11 | ... | ... | ... | ... | ... | ... | ... | ... | ... | 19 | 3.895 | 14 | 82 | 3.563 | 12 |
| 4200.10 | N III | 6 | ... | ... | ... | ... | ... | ... | ... | ... | ... | * | * | * | * | * | * |
| 4227.20 | [Fe V] | | ... | ... | ... | 11 | 0.134 | 14 | 50 | 0.152 | 20 | 40 | 3.517 | 16 | 106 | 3.462 | 12 |
| 4236.91 | N II | 48 | 22 | 0.051 | 38 | ... | ... | ... | ... | ... | ... | ... | ... | ... | ... | ... | ... |
| 4237.05 | N II | 48 | * | * | * | ... | ... | ... | ... | ... | ... | ... | ... | ... | ... | ... | ... |
| 4241.78 | N II | 48 | 12 | 0.077 | 28 | ... | ... | ... | ... | ... | ... | ... | ... | ... | ... | ... | ... |
| 4267.15 | C II | 6 | 14 | 0.781 | 7 | -15 | 0.814 | 6 | 34 | 1.246 | 6 | -2 | 0.487 | : | 84 | 1.029 | 30 |
| 4303.61 | O II | 65 | 29 | 0.086 | 25 | ... | ... | ... | ... | ... | ... | ... | ... | ... | ... | ... | ... |
| 4303.82 | O II | 53a | * | * | * | ... | ... | ... | ... | ... | ... | ... | ... | ... | ... | ... | ... |
| 4317.14 | O II | 2 | 15 | 0.210[g] | 13 | ... | ... | ... | ... | ... | ... | ... | ... | ... | ... | ... | ... |
| 4319.63 | O II | 2 | 17 | 0.081 | 26 | ... | ... | ... | ... | ... | ... | ... | ... | ... | ... | ... | ... |
| 4325.76 | O II | 2 | 16 | 0.036 | : | ... | ... | ... | ... | ... | ... | ... | ... | ... | ... | ... | ... |
| 4336.83 | O II | 2 | 19 | 0.054 | 36 | ... | ... | ... | ... | ... | ... | ... | ... | ... | ... | ... | ... |
| 4338.67 | He II | 4.10 | ... | ... | ... | -10 | 0.919 | 6 | 30 | 1.066 | 6 | 22 | 4.806 | 12 | 82 | 4.090 | 10 |
| 4340.47 | H I | H5 | 12 | 45.666 | 5 | -17 | 46.576 | 5 | 31 | 48.072 | 5 | 19 | 47.240 | 5 | 80 | 47.029 | 5 |
| 4345.55 | O II | 65c | 15 | 0.257[g] | 11 | ... | ... | ... | ... | ... | ... | ... | ... | ... | ... | ... | ... |
| 4345.56 | O II | 2 | * | * | * | -14 | 0.020 | : | 32 | 0.038 | : | ... | ... | ... | ... | ... | ... |
| 4349.43 | O II | 2 | 14 | 0.197 | 13 | -16 | 0.048 | 31 | 31 | 0.044 | : | ... | ... | ... | ... | ... | ... |
| 4363.21 | [O III] | 2F | 12 | 0.528 | 7 | -16 | 14.546 | 5 | 30 | 13.242 | 5 | 17 | 18.793 | 6 | 83 | 22.773 | 5 |
| 4366.89 | O II | 2 | 12 | 0.199 | 13 | ... | ... | ... | ... | ... | ... | ... | ... | ... | ... | ... | ... |
| 4387.93 | He I | 51 | 13 | 1.101[e] | 6 | -18 | 1.512[e] | 5 | 36 | 1.833[e] | 6 | ... | ... | ... | ... | ... | ... |
| 4391.94 | Ne II | 57 | 20 | 0.062 | 32 | ... | ... | ... | ... | ... | ... | ... | ... | ... | ... | ... | ... |
| 4409.30 | Ne II | 57 | 15 | 0.061 | 32 | ... | ... | ... | ... | ... | ... | ... | ... | ... | ... | ... | ... |
| 4414.90 | O II | 5 | 17 | 0.036 | : | ... | ... | ... | ... | ... | ... | ... | ... | ... | ... | ... | ... |
| 4416.97 | O II | 5 | 15 | 0.090 | 24 | ... | ... | ... | ... | ... | ... | ... | ... | ... | ... | ... | ... |
| 4437.55 | He I | 50 | 14 | 0.092 | 23 | -19 | 0.061 | 25 | 32 | 0.064 | : | ... | ... | ... | ... | ... | ... |
| 4465.41 | O II | 94 | 19 | 0.030 | : | ... | ... | ... | ... | ... | ... | ... | ... | ... | ... | ... | ... |
| 4467.92 | O II | 94 | 5 | 0.018 | : | ... | ... | ... | ... | ... | ... | ... | ... | ... | ... | ... | ... |
| 4471.47 | He I | 14 | 15 | 6.476 | 5 | -17 | 3.840 | 5 | 35 | 4.267 | 5 | 18 | 1.609 | 28 | 88 | 2.085 | 16 |
| 4491.14 | [Fe IV] | | ... | ... | ... | -9 | 0.027 | : | 39 | 0.052 | : | ... | ... | ... | ... | ... | ... |

**Table 3.** continued.

| | | | PB8 | | | NGC 2867–1 | | | NGC 2867–2 | | | PB6–1 | | | PB6–2 | | |
|---|---|---|---|---|---|---|---|---|---|---|---|---|---|---|---|---|---|
| $\lambda_0$ (Å) | Ion | Mult. | $V_{rad}$ | $I(\lambda)$ [a] | Err(%) | $V_{rad}$ | $I(\lambda)$ [b] | Err(%) | $V_{rad}$ | $I(\lambda)$ [b] | Err(%) | $V_{rad}$ | $I(\lambda)$ [c] | Err(%) | $V_{rad}$ | $I(\lambda)$ [c] | Err(%) |
| 4541.59 | He II | 4.9 | ... | ... | ... | -10 | 1.384 | 5 | 28 | 1.343 | 6 | 22 | 7.146 | 9 | 81 | 6.568 | 8 |
| 4562.60 | Mg I] | 1 | ... | ... | ... | -25 | 0.187 | 10 | 37 | 0.285 | 12 | ... | ... | ... | ... | ... | ... |
| 4571.10 | Mg I] | 1 | ... | ... | ... | -24 | 0.513 | 6 | 35 | 0.638 | 7 | ... | ... | ... | ... | ... | ... |
| 4590.97 | O II | 15 | 16 | 0.066 | 29 | ... | ... | ... | ... | ... | ... | ... | ... | ... | ... | ... | ... |
| 4595.95 | O II | 15 | 30 | 0.044 | : | ... | ... | ... | ... | ... | ... | ... | ... | ... | ... | ... | ... |
| 4596.18 | O II | 15 | * | * | * | ... | ... | ... | ... | ... | ... | ... | ... | ... | ... | ... | ... |
| 4601.48 | N II | 5 | 14 | 0.099 | 21 | ... | ... | ... | ... | ... | ... | ... | ... | ... | ... | ... | ... |
| 4607.16 | N II | 5 | 10 | 0.083 | 25 | ... | ... | ... | ... | ... | ... | ... | ... | ... | ... | ... | ... |
| 4609.44 | O II | 93 | 10 | 0.033 | : | ... | ... | ... | ... | ... | ... | ... | ... | ... | ... | ... | ... |
| 4613.87 | N II | 5 | 12 | 0.063 | 30 | ... | ... | ... | ... | ... | ... | ... | ... | ... | ... | ... | ... |
| 4621.39 | N II | 5 | 11 | 0.085 | 24 | ... | ... | ... | ... | ... | ... | ... | ... | ... | ... | ... | ... |
| 4630.54 | N II | 5 | 13 | 0.289 | 10 | ... | ... | ... | ... | ... | ... | ... | ... | ... | ... | ... | ... |
| 4634.14 | N III | 2 | 16 | 0.141 | 16 | -13 | 0.704 | 6 | 28 | 0.501 | 8 | 19 | 1.072 | 38 | 81 | 1.147 | 25 |
| 4638.86 | O II | 1 | 12 | 0.206 | 12 | -17 | 0.071 | 21 | 30 | 0.087 | 29 | ... | ... | ... | ... | ... | ... |
| 4640.64 | N III | 2 | 14 | 0.234 | 11 | -14 | 1.552 | 5 | 27 | 1.155 | 6 | 16 | 1.603 | 27 | 82 | 2.268 | 15 |
| 4641.81 | O II | 1 | 14 | 0.380 | 8 | -10 | 0.281 | 8 | 32 | 0.263 | 12 | ... | ... | ... | ... | ... | ... |
| 4641.85 | N III | 2 | * | * | * | * | * | * | * | * | * | ... | ... | ... | ... | ... | ... |
| 4643.06 | N II | 5 | 14 | 0.122 | 18 | ... | ... | ... | ... | ... | ... | ... | ... | ... | ... | ... | ... |
| 4647.42 | C III | 1 | ... | ... | ... | -12 | 0.342 | 7 | 28 | 0.366 | 10 | 22 | 1.153 | 36 | 83 | 0.636 | : |
| 4649.13 | O II | 1 | 14 | 0.458 | 8 | -16 | 0.147 | 12 | 32 | 0.215 | 14 | ... | ... | ... | ... | ... | ... |
| 4650.25 | C III | 1 | ... | ... | ... | -12 | 0.212 | 9 | 26 | 0.339 | 10 | 17 | 0.734 | : | 82 | 0.553 | : |
| 4650.84 | O II | 1 | 12 | 0.221 | 12 | -12 | 0.339 | 7 | 29 | 0.193 | 15 | ... | ... | ... | ... | ... | ... |
| 4651.47 | C III | 1 | ... | ... | ... | -12 | 0.109 | 15 | 27 | 0.097 | 27 | ... | ... | ... | ... | ... | ... |
| 4657.56 | C IV | | ... | ... | ... | -7 | 0.083 | : | 17 | 0.245 | : | ... | ... | ... | ... | ... | ... |
| 4658.10 | [Fe III] | 3F | 15 | 0.131[e] | 17 | 14 | 0.171 | 11 | 42 | 0.404 | 9 | 41 | 3.136 | 16 | 103 | 1.595 | 19 |
| 4661.63 | O II | 1 | 14 | 0.222 | 12 | -17 | 0.062 | 23 | 31 | 0.067 | 36 | ... | ... | ... | ... | ... | ... |
| 4673.73 | O II | 1 | 13 | 0.037 | : | ... | ... | ... | ... | ... | ... | ... | ... | ... | ... | ... | ... |
| 4676.24 | O II | 1 | 19 | 0.184[e] | 13 | ... | ... | ... | ... | ... | ... | ... | ... | ... | ... | ... | ... |
| 4685.68 | He II | 3.4 | ... | ... | ... | -6 | 38.149 | 5 | 32 | 34.425 | 5 | 24 | 180.051 | 5 | 84 | 156.554 | 5 |
| 4699.22 | O II | 25 | 11 | 0.026 | : | ... | ... | ... | ... | ... | ... | ... | ... | ... | ... | ... | ... |
| 4701.62 | [Fe III] | 3F | 6 | 0.039 | : | -29 | 0.036 | 37 | 34 | 0.085 | 29 | ... | ... | ... | ... | ... | ... |
| 4711.37 | [Ar IV] | 1F | 15 | 0.078 | 25 | -13 | 2.547 | 5 | 27 | 1.816 | 5 | 20 | 13.354 | 6 | 78 | 12.257 | 6 |
| 4713.14 | He I | 12 | 13 | 0.624 | 7 | -18 | 0.602 | 6 | 33 | 0.569 | 8 | ... | ... | ... | ... | ... | ... |
| 4714.36 | [Ne IV] | | ... | ... | ... | -20 | 0.156 | 11 | 18 | 0.136 | 20 | 12 | 2.859 | 17 | 72 | 2.316[j] | 14 |
| 4715.80 | [Ne IV] | | ... | ... | ... | -18 | 0.040 | 33 | 17 | 0.038 | : | 10 | 1.466[j] | 29 | 72 | 0.731 | 35 |
| 4724.15 | [Ne IV] | 1F | ... | ... | ... | -11 | 0.156 | 11 | 25 | 0.139 | 20 | 18 | 2.881 | 16 | 81 | 2.897[j] | 12 |
| 4725.62 | [Ne IV] | 1F | ... | ... | ... | -13 | 0.128 | 13 | 23 | 0.100 | 26 | 18 | 2.516[j] | 18 | 77 | 2.275 | 14 |
| 4740.17 | [Ar IV] | 1F | 17 | 0.065[e] | 29 | -10 | 3.368[e] | 5 | 29 | 1.971 | 5 | 22 | 11.856 | 7 | 81 | 10.967 | 6 |
| 4788.13 | N II | 20 | 10 | 0.067 | 28 | ... | ... | ... | ... | ... | ... | ... | ... | ... | ... | ... | ... |
| 4803.29 | N II | 20 | ... | ...[e] | ... | ... | ... | ... | ... | ... | ... | ... | ... | ... | ... | ... | ... |
| 4859.32 | He II | 4.8 | ... | ... | ... | -11 | 1.855 | 5 | 28 | 2.135 | 5 | 20 | 8.851 | 8 | 79 | 8.473 | 7 |
| 4861.33 | H I | H4 | 10 | 100.013 | 5 | -19 | 100.024 | 5 | 30 | 100.008 | 5 | 17 | 100.014 | 5 | 78 | 99.986 | 5 |
| 4881.00 | [Fe III] | 2F | 5 | 0.023 | : | -25 | 0.042 | 31 | 36 | 0.085 | 28 | ... | ... | ... | ... | ... | ... |
| 4890.86 | O II | 28 | 10 | 0.121[e] | 18 | ... | ... | ... | ... | ... | ... | ... | ... | ... | ... | ... | ... |
| 4906.81 | O II | 28 | 12 | 0.096 | 21 | ... | ... | ... | ... | ... | ... | ... | ... | ... | ... | ... | ... |
| 4921.93 | He I | 48 | 11 | 1.737 | 5 | -21 | 0.955 | 5 | 31 | 1.077 | 6 | ... | ... | ... | ... | ... | ... |
| 4924.53 | O II | 28 | 12 | 0.154 | 15 | ... | ... | ... | ... | ... | ... | ... | ... | ... | ... | ... | ... |

**Table 3.** continued.

| | | | PB8 | | | NGC 2867–1 | | | NGC 2867–2 | | | PB6–1 | | | PB6–2 | | |
|---|---|---|---|---|---|---|---|---|---|---|---|---|---|---|---|---|---|
| $\lambda_0$ (Å) | Ion | Mult. | $V_{rad}$ | $I(\lambda)$ [a] | Err(%) | $V_{rad}$ | $I(\lambda)$ [b] | Err(%) | $V_{rad}$ | $I(\lambda)$ [b] | Err(%) | $V_{rad}$ | $I(\lambda)$ [c] | Err(%) | $V_{rad}$ | $I(\lambda)$ [c] | Err(%) |
| 4931.32 | [O III] | 1F | ... | ... | ... | -23 | 0.151 | 11 | 26 | 0.157 | 17 | ... | ... | ... | ... | ... | ... |
| 4958.91 | [O III] | 1F | 10 | 116.957 | 5 | -20 | 432.228 | 5 | 29 | 421.200 | 5 | 14 | 286.449 | 5 | 82 | 368.783 | 5 |
| 4994.37 | N II | 24 | 10 | 0.099 | 21 | ... | ... | ... | ... | ... | ... | ... | ... | ... | ... | ... | ... |
| 5001.13 | N II | 19 | 22 | 0.155 | 15 | ... | ... | ... | ... | ... | ... | ... | ... | ... | ... | ... | ... |
| 5001.47 | N II | 19 | * | * | * | ... | ... | ... | ... | ... | ... | ... | ... | ... | ... | ... | ... |
| 5006.84 | [O III] | 1F | ... | 348.532[i] | ... | -20 | 1199.61 | 5 | 29 | 1308.389 | 5 | 14 | 820.993 | 5 | 81 | 1074.22 | 5 |
| 5015.68 | He I | 4 | 8 | 2.657 | 5 | -23 | 1.421 | 5 | 30 | 1.739 | 5 | 20 | 0.985 | 38 | 79 | 0.732 | 33 |
| 5047.74 | He I | 47 | 7 | 0.231 | 23 | -24 | 0.145 | 11 | 28 | 0.246 | 12 | ... | ... | ... | ... | ... | ... |
| 5146.80 | [Fe VI] | ... | ... | ... | ... | ... | ... | ... | ... | ... | ... | 18 | 2.814 | 13 | 77 | 2.016 | 12 |
| 5191.82 | [Ar III] | 3F | ... | ... | ... | -31 | 0.127 | 12 | 20 | 0.153 | 17 | ... | ... | ... | ... | ... | ... |
| 5197.90 | [N I] | 1F | ... | ... | ... | -34 | 0.301 | 7 | 31 | 0.503 | 8 | ... | ... | ... | ... | ... | ... |
| 5200.26 | [N I] | 1F | ... | ... | ... | -35 | 0.199 | 9 | 31 | 0.365 | 9 | ... | ... | ... | ... | ... | ... |
| 5346.10 | [Kr IV] | ... | ... | ... | ... | -28 | 0.128 | 12 | 15 | 0.155 | 16 | ... | ... | ... | ... | ... | ... |
| 5411.52 | He II | 4.7 | ... | ... | ... | -18 | 2.960 | 5 | 20 | 3.175 | 5 | 12 | 14.980 | 6 | 72 | 13.799 | 5 |
| 5517.71 | [Cl III] | 1F | 2 | 0.366 | 14 | -28 | 0.583 | 6 | 21 | 0.670 | 7 | ... | ... | ... | ... | ... | ... |
| 5537.88 | [Cl III] | 1F | -1 | 0.366 | 14 | -29 | 0.718 | 6 | 20 | 0.828 | 6 | ... | ... | ... | ... | ... | ... |
| 5592.37 | O III | 5 | ... | ... | ... | -23 | 0.082 | 15 | 14 | 0.156 | 16 | ... | ... | ... | ... | ... | ... |
| 5666.64 | N II | 3 | 2 | 0.192 | 25 | ... | ... | ... | ... | ... | ... | ... | ... | ... | ... | ... | ... |
| 5676.02 | N II | 3 | 0 | 0.084 | : | ... | ... | ... | ... | ... | ... | ... | ... | ... | ... | ... | ... |
| 5677.00 | [Fe VI] | ... | ... | ... | ... | ... | ... | ... | ... | ... | ... | -1 | 1.863 | 17 | 72 | 0.740 | 25 |
| 5679.56 | N II | 3 | 4 | 0.260 | 18 | ... | ... | ... | ... | ... | ... | ... | ... | ... | ... | ... | ... |
| 5754.64 | [N II] | 3F | -1 | 0.346 | 14 | -38 | 1.159 | 5 | 24 | 1.504 | 6 | 3 | 2.324 | 14 | 74 | 4.245 | 7 |
| 5801.33 | C IV | 1 | ... | ... | ... | -17 | 0.164 | 10 | 18 | 0.144 | 16 | ... | ... | ... | ... | ... | ... |
| 5811.98 | C IV | 1 | ... | ... | ... | -18 | 0.079 | 16 | 18 | 0.115 | 19 | ... | ... | ... | ... | ... | ... |
| 5820.40 | He II | 5.34 | ... | ... | ... | -13 | 0.018 | : | 16 | 0.014 | : | ... | ... | ... | ... | ... | ... |
| 5828.60 | He II | 5.33 | ... | ... | ... | -31 | 0.020 | : | 8 | 0.018 | : | ... | ... | ... | ... | ... | ... |
| 5868.00 | [Kr IV] | ... | ... | ... | ... | -35 | 0.216 | 8 | 6 | 0.297 | 10 | ... | ... | ... | ... | ... | ... |
| 5875.64 | He I | 11 | 5 | 17.127 | 6 | -27 | 10.919 | 5 | 24 | 13.120 | 5 | 7 | 4.066 | 9 | 75 | 5.245 | 6 |
| 5931.78 | N II | 28 | 10 | 0.151 | 30 | ... | ... | ... | ... | ... | ... | ... | ... | ... | ... | ... | ... |
| 5941.65 | N II | 28 | 6 | 0.115 | : | ... | ... | ... | ... | ... | ... | ... | ... | ... | ... | ... | ... |
| 5977.03 | He II | 5.23 | ... | ... | ... | -22 | 0.045 | 25 | 9 | 0.126 | 17 | ... | ... | ... | ... | ... | ... |
| 6004.73 | He II | 5.22 | ... | ... | ... | -20 | 0.058 | 20 | 17 | 0.097 | 21 | ... | ... | ... | ... | ... | ... |
| 6036.70 | He II | 5.21 | ... | ... | ... | -14 | 0.058 | 20 | 21 | 0.071 | 27 | ... | ... | ... | ... | ... | ... |
| 6074.10 | He II | 5.20 | ... | ... | ... | -12 | 0.075 | 16 | 24 | 0.076 | 26 | ... | ... | ... | ... | ... | ... |
| 6101.83 | [K IV] | 1F | ... | ... | ... | -24 | 0.190 | 9 | 17 | 0.164 | 14 | ... | ... | ... | ... | ... | ... |
| 6118.20 | He II | 5.19 | ... | ... | ... | -14 | 0.080 | 16 | 24 | 0.091 | 22 | ... | ... | ... | ... | ... | ... |
| 6170.60 | He II | 5.18 | ... | ... | ... | -16 | 0.111 | 12 | 28 | 0.124 | 17 | ... | ... | ... | ... | ... | ... |
| 6233.80 | He II | 5.27 | ... | ... | ... | -15 | 0.122 | 11 | 22 | 0.115 | 18 | 11 | 0.857 | 32 | 74 | 0.841 | 20 |
| 6300.30 | [O I] | 1F | ... | ... | ... | -33 | 6.095 | 5 | 30 | 8.846 | 6 | 3 | 1.491 | 19 | 82 | 4.177 | 7 |
| 6312.10 | [S III] | 3F | 3 | 0.639 | 9 | -26 | 1.942 | 5 | 23 | 1.991 | 6 | 9 | 4.169 | 9 | 73 | 3.552 | 7 |
| 6347.11 | Si II | 2 | 4 | 0.098 | : | ... | ... | ... | ... | ... | ... | ... | ... | ... | ... | ... | ... |
| 6363.78 | [O I] | 1F | ... | ... | ... | -33 | 1.965 | 5 | 30 | 2.857 | 6 | 4 | 0.505 | : | 81 | 1.395 | 13 |
| 6371.36 | Si II | 2 | 4 | 0.066 | : | -27 | 0.029 | 34 | 25 | 0.031 | : | ... | ... | ... | ... | ... | ... |
| 6406.30 | He II | 5.15 | ... | ... | ... | -13 | 0.181 | 9 | 26 | 0.184 | 13 | 22 | 0.493 | : | 79 | 0.522 | 30 |
| 6435.10 | [Ar V] | ... | ... | ... | ... | -20 | 0.125 | 11 | 13 | 0.108 | 19 | 14 | 2.836 | 11 | 65 | 2.281 | 9 |
| 6461.95 | C II | 17.04 | ... | ... | ... | -26 | 0.074 | 16 | 20 | 0.108 | 19 | ... | ... | ... | ... | ... | ... |
| 6527.11 | He II | 5.14 | ... | ... | ... | -17 | 0.209 | 8 | 21 | 0.201 | 12 | 17 | 0.695 | 38 | 73 | 1.001 | 16 |

| | | | PB8 | | | NGC 2867–1 | | | NGC 2867–2 | | | PB6–1 | | | PB6–2 | | |
|---|---|---|---|---|---|---|---|---|---|---|---|---|---|---|---|---|---|
| $\lambda_0$ (Å) | Ion | Mult. | $V_{rad}$ | $I(\lambda)^a$ | Err(%) | $V_{rad}$ | $I(\lambda)^b$ | Err(%) | $V_{rad}$ | $I(\lambda)^b$ | Err(%) | $V_{rad}$ | $I(\lambda)^c$ | Err(%) | $V_{rad}$ | $I(\lambda)^c$ | Err(%) |
| 6548.03 | [N II] | 1F | 3 | 7.667 | 6 | -31 | 17.613 | 5 | 32 | 24.273 | 6 | 9 | 34.261 | 6 | 81 | 57.700 | 5 |
| 6560.00 | He II | 4.6 | ... | ... | ... | -11 | 5.556 | 5 | 29 | 6.672 | 6 | 22 | 26.517 | 6 | 79 | 22.001 | 6 |
| 6562.82 | H I | H3 | 6 | 282.564 | 6 | -23 | 282.343 | 5 | 26 | 288.854 | 6 | 12 | 279.134 | 6 | 72 | 278.862 | 5 |
| 6578.05 | C II | 2 | 8 | 0.545 | 9 | -22 | 0.302 | 7 | 26 | 0.598 | 7 | ... | ... | ... | ... | ... | ... |
| 6583.41 | [N II] | 1F | 3 | 22.318 | 6 | -31 | 51.293 | 5 | 32 | 69.372 | 6 | 9 | 100.546 | 6 | 81 | 164.419 | 5 |
| 6678.15 | He I | 46 | 7 | 5.233 | 6 | -24 | 3.129 | 6 | 28 | 3.643 | 6 | 9 | 1.079 | 24 | 77 | 1.502 | 12 |
| 6683.20 | He II | 5.13 | ... | ... | ... | -15 | 0.289 | 7 | 22 | 0.360 | 9 | 15 | 1.294 | 21 | 74 | 1.220 | 14 |
| 6716.47 | [S II] | 2F | 2 | 0.957 | 7 | -32 | 3.658 | 6 | 31 | 6.201 | 6 | 8 | 3.041 | 10 | 79 | 4.327 | 7 |
| 6730.85 | [S II] | 2F | 1 | 1.441 | 7 | -32 | 5.975 | 6 | 31 | 9.245 | 6 | 9 | 4.064 | 9 | 78 | 6.362 | 6 |
| 6779.93 | C II | 14 | ... | ... | ... | -20 | 0.045 | 23 | 21 | 0.079 | 23 | ... | ... | ... | ... | ... | ... |
| 6795.00 | [K IV] | 1F | ... | ... | ... | -17 | 0.049 | 21 | 23 | 0.066 | 27 | ... | ... | ... | ... | ... | ... |
| 6890.88 | He II | 5.12 | ... | ... | ... | -13 | 0.353 | 7 | 23 | 0.326 | 9 | 16 | 1.100 | 23 | 72 | 1.116 | 14 |
| 7005.67 | [Ar V] | ... | ... | ... | ... | -12 | 0.242 | 8 | 21 | 0.150 | 14 | 21 | 6.103 | 8 | 72 | 4.877 | 7 |
| 7062.26 | He I | 1/11 | ... | ... | ... | -12 | 0.060 | 18 | 28 | 0.049 | 33 | ... | ... | ... | ... | ... | ... |
| 7065.28 | He I | 10 | 6 | 4.265 | 7 | -25 | 4.127 | 6 | 25 | 4.611 | 6 | 7 | 1.269 | 20 | 76 | 1.770 | 10 |
| 7135.78 | [Ar III] | 1F | 8 | 15.477 | 7 | -21 | 13.628 | 6 | 27 | 14.757 | 6 | 11 | 14.155 | 7 | 76 | 16.249 | 6 |
| 7170.62 | [Ar IV] | 2F | ... | ... | ... | -15 | 0.118 | 11 | 24 | 0.106 | 18 | 19 | 0.776 | 32 | 77 | 0.603 | 24 |
| 7177.50 | He II | 5.11 | ... | ... | ... | -12 | 0.408 | 7 | 25 | 0.448 | 8 | 16 | 2.239 | 13 | 76 | 2.028 | 9 |
| 7231.34 | C II | 3 | 7 | 0.234 | 17 | -22 | 0.138 | 10 | 25 | 0.287 | 9 | ... | ... | ... | ... | ... | ... |
| 7236.42 | C II | 3 | 11 | 0.464 | 10 | -19 | 0.266 | 7 | 29 | 0.800 | 7 | ... | ... | ... | ... | ... | ... |
| 7262.76 | [Ar IV] | 2F | ... | ... | ... | -8 | 0.101 | 12 | 30 | 0.089 | 20 | 23 | 0.760 | 32 | 77 | 0.414 | 33 |
| 7281.35 | He I | 45 | 8 | 0.815 | 8 | -23 | 0.614 | 6 | 28 | 0.666 | 7 | ... | ... | ... | ... | ... | ... |
| 7318.92 | [O II] | 2F | 9 | 0.227 | 18 | ... | ...$^j$ | ... | ... | ...$^j$ | ... | ... | ...$^j$ | ... | ... | ...$^j$ | ... |
| 7319.99 | [O II] | 2F | 11 | 0.811 | 8 | ... | ...$^j$ | ... | ... | ...$^j$ | ... | ... | ...$^j$ | ... | ... | ...$^j$ | ... |
| 7329.66 | [O II] | 2F | 5 | 0.387 | 12 | ... | ...$^j$ | ... | ... | ...$^j$ | ... | ... | ...$^j$ | ... | ... | ...$^j$ | ... |
| 7330.73 | [O II] | 2F | 6 | 0.471 | 10 | ... | ...$^j$ | ... | ... | ...$^j$ | ... | ... | ...$^j$ | ... | ... | ...$^j$ | ... |
| 7377.83 | [Ni II] | ... | ... | ... | ... | -26 | 0.015 | : | 36 | 0.060 | 27 | ... | ... | ... | ... | ... | ... |
| 7499.85 | He I | 1/8 | ... | ... | ... | -21 | 0.033 | 27 | 31 | 0.046 | 37 | ... | ... | ... | ... | ... | ... |
| 7530.54 | [Cl IV] | ... | ... | ... | ... | -19 | 0.263 | 7 | 20 | 0.193 | 11 | 14 | 0.554 | : | 74 | 0.580 | 26 |
| 7535.40 | [Xe IV] | ... | ... | ... | ... | -35 | 0.040 | 22 | 8 | 0.068 | 26 | ... | ... | ... | ... | ... | ... |
| 7592.74 | He II | 5.10 | ... | ... | ... | -15 | 0.555 | 6 | 23 | 0.252 | 9 | 13 | 1.967 | 14 | 76 | 1.222 | 15 |
| 7726.20 | C IV | 8.01 | ... | ... | ... | -19 | 0.133 | 10 | 20 | 0.077 | 19 | 11 | 0.954 | 20 | 73 | 0.678 | 19 |
| 7751.10 | [Ar III] | 2F | 7 | 3.493 | 7 | -24 | 3.123 | 6 | 26 | 3.274 | 7 | 10 | 3.327 | 9 | 74 | 3.945 | 7 |
| 7816.13 | He I | 1/7 | 8 | 0.076 | : | -24 | 0.042 | 22 | 26 | 0.052 | 33 | ... | ... | ... | ... | ... | ... |
| 8045.63 | [Cl IV] | 1F | ... | ... | ... | -13 | 0.534 | 6 | 26 | 0.457 | 8 | 18 | 1.541 | 16 | 77 | 1.729 | 10 |
| 8196.48 | C III | 43 | ... | ... | ... | -11 | 0.247 | 7 | 30 | 0.195 | 11 | ... | ... | ... | ... | ... | ... |
| 8236.77 | He II | 5.9 | ... | ... | ... | -15 | 0.787 | 6 | 24 | 0.785 | 7 | 16 | 3.685 | 10 | 74 | 3.655 | 7 |
| 8247.73 | H I | P41 | 9 | 0.079 | : | ... | ... | ... | ... | ... | ... | ... | ... | ... | ... | ... | ... |
| 8249.97 | H I | P40 | 4 | 0.061 | : | ... | ... | ... | ... | ... | ... | ... | ... | ... | ... | ... | ... |
| 8252.40 | H I | P39 | 7 | 0.077 | : | ... | ... | ... | ... | ... | ... | ... | ... | ... | ... | ... | ... |
| 8255.02 | H I | P38 | 6 | 0.048 | : | ... | ... | ... | ... | ... | ... | ... | ... | ... | ... | ... | ... |
| 8257.85 | H I | P37 | 10 | 0.091 | : | ... | ... | ... | ... | ... | ... | ... | ... | ... | ... | ... | ... |
| 8260.93 | H I | P36 | 9 | 0.087 | : | ... | ... | ... | ... | ... | ... | ... | ... | ... | ... | ... | ... |
| 8264.28 | H I | P35 | 17 | 0.113 | 36 | ... | ... | ... | ... | ... | ... | ... | ... | ... | ... | ... | ... |
| 8267.94 | H I | P34 | 5 | 0.075 | : | ... | ... | ... | ... | ... | ... | ... | ... | ... | ... | ... | ... |
| 8271.93 | H I | P33 | 11 | 0.088 | : | ... | ... | ... | ... | ... | ... | ... | ... | ... | ... | ... | ... |
| 8286.43 | H I | P30 | -3 | 0.101 | 28 | ... | ... | ... | ... | ... | ... | ... | ... | ... | ... | ... | ... |

**Table 3.** continued.

| | | | PB8 | | | NGC 2867–1 | | | NGC 2867–2 | | | PB6–1 | | | PB6–2 | | |
|---|---|---|---|---|---|---|---|---|---|---|---|---|---|---|---|---|---|
| $\lambda_0$ (Å) | Ion | Mult. | $V_{rad}$ | $I(\lambda)$ [a] | Err(%) | $V_{rad}$ | $I(\lambda)$ [b] | Err(%) | $V_{rad}$ | $I(\lambda)$ [b] | Err(%) | $V_{rad}$ | $I(\lambda)$ [c] | Err(%) | $V_{rad}$ | $I(\lambda)$ [c] | Err(%) |
| 8292.31 | H I | P29 | 7 | 0.130 | 29 | -23 | 0.091 | 12 | 23 | 0.101 | 16 | ... | ... | ... | ... | ... | ... |
| 8306.11 | H I | P27 | 6 | 0.120 | 27 | -26 | 0.115 | 10 | 24 | 0.150 | 12 | ... | ... | ... | ... | ... | ... |
| 8314.26 | H I | P26 | 5 | 0.133 | 23 | -28 | 0.145 | 9 | 23 | 0.146 | 12 | ... | ... | ... | ... | ... | ... |
| 8323.42 | H I | P25 | 6 | 0.159 | 23 | -25 | 0.149 | 9 | 25 | 0.145 | 13 | ... | ... | ... | ... | ... | ... |
| 8333.78 | H I | P24 | 8 | 0.165 | 29 | -25 | 0.195 | 9 | 25 | 0.147 | 15 | ... | ... | ... | ... | ... | ... |
| 8345.55 | H I | P23 | 6 | 0.185 | 22 | -24 | 0.144 | 10 | 22 | 0.186 | 13 | ... | ... | ... | ... | ... | ... |
| 8359.00 | H I | P22 | 7 | 0.192 | 19 | -25 | 0.190 | 8 | 24 | 0.209 | 11 | ... | ... | ... | ... | ... | ... |
| 8361.67 | He I | 1/6 | 10 | 0.104 | 35 | -27 | 0.067 | 15 | 27 | 0.093 | 18 | ... | ... | ... | ... | ... | ... |
| 8374.48 | H I | P21 | 7 | 0.220 | 16 | -24 | 0.207 | 8 | 25 | 0.244 | 10 | ... | ... | ... | ... | ... | ... |
| 8392.40 | H I | P20 | 6 | 0.244 | 15 | -24 | 0.228 | 8 | 24 | 0.269 | 10 | ... | ... | ... | ... | ... | ... |
| 8413.32 | H I | P19 | 5 | 0.272 | 15 | -23 | 0.265 | 7 | 22 | 0.217 | 11 | ... | ... | ... | ... | ... | ... |
| 8421.80 | He I | 6/18 | 6 | 0.057 | : | ... | ... | ... | ... | ... | ... | ... | ... | ... | ... | ... | ... |
| 8437.96 | H I | P18 | 5 | 0.326 | 12 | -23 | 0.310 | 7 | 25 | 0.348 | 9 | ... | ... | ... | ... | ... | ... |
| 8467.25 | H I | P17 | 5 | 0.393 | 11 | -23 | 0.348 | 7 | 24 | 0.364 | 9 | ... | ... | ... | ... | ... | ... |
| 8480.90 | [Cl III] | 3F | ... | ... | ... | -28 | 0.025 | : | 18 | 0.041 | : | ... | ... | ... | ... | ... | ... |
| 8502.48 | H I | P16 | 6 | 0.464 | 10 | -25 | 0.421 | 7 | 23 | 0.362 | 9 | ... | ... | ... | ... | ... | ... |
| 8545.38 | H I | P15 | 6 | 0.538 | 11 | -26 | 0.484 | 7 | 22 | 0.526 | 8 | ... | ... | ... | ... | ... | ... |
| 8598.39 | H I | P14 | 6 | 0.661 | 9 | -24 | 0.606 | 7 | 24 | 0.645 | 8 | ... | ... | ... | ... | ... | ... |
| 8665.02 | H I | P13 | 5 | 0.828 | 9 | -23 | 0.802 | 6 | 24 | 0.794 | 8 | ... | ... | ... | ... | ... | ... |
| 8733.43 | He I | 6/12 | 6 | 0.069 | : | ... | ... | ... | ... | ... | ... | ... | ... | ... | ... | ... | ... |
| 8750.47 | H I | P12 | 4 | 1.015 | 9 | -28 | 0.955 | 6 | 21 | 0.998 | 7 | 7 | 0.907 | 20 | 62 | 1.068 | 12 |
| 8845.38 | He I | 6/11 | 5 | 0.114 | 33 | -28 | 0.049 | ... | | 0.088 | ... | ... | ... | ... | ... | ... | ... |
| 8862.79 | H I | P11 | 4 | 1.196 | 9 | -25 | 1.175 | 7 | 23 | 1.245 | 8 | 9 | 1.183 | 16 | 71 | 1.070 | 12 |
| 8996.99 | He I | 6/10 | 4 | 0.133 | : | -33 | 0.068 | ... | | 0.100 | ... | ... | ... | ... | ... | ... | ... |

[a] Where $I$ is the reddened corrected flux, with c(H$\beta$)=0.36, in units of 100.00 = $3.384 \times 10^{-13}$.
[b] Where $I$ is the reddened corrected flux, with c(H$\beta$)=0.39 and 0.43 dex, in units of 100.00 = $2.984 \times 10^{-13}$
and 100.00 = $1.853 \times 10^{-13}$ergs cm$^{-2}$ s$^{-1}$ for components 1 ad 2 respectively.
[c] Where $I$ is the reddened corrected flux, with c(H$\beta$)=0.47 and 0.53 dex, in units of 100.00 = $1.692 \times 10^{-14}$
and 100.00 = $3.161 \times 10^{-14}$ergs cm$^{-2}$ s$^{-1}$ for components 1 ad 2 respectively.
[d] Deblended with DIPSO.
[e] Affected by ghosts.
[f] Affected by telluric emission.
[g] Affected by un unknown emission.
[h] Partially blended with H6 from the other component.
[i] Saturated in long and short exposures. Intensity computed from the theoretical ratio $I(\lambda 5007)/I(\lambda 4958)$=2.98
[j] Blend of different components

**Table 4.** Extinction coefficients.

| Object | c(Hβ) | E(B−V) | Lines used | $T_e$ (K), $n_e$ (cm$^{-3}$) |
|---|---|---|---|---|
| PB 8 | 0.36±0.05 | 0.24 | H16 to H3 and P22 to P11 | 6500, 4000 |
| NGC 2867–1 | 0.39±0.03 | 0.26 | H16 to H3 and P22 to P11 | 10000, 2600 |
| NGC 2867–2 | 0.43±0.04 | 0.29 | H16 to H3 and P22 to P11 | 10000, 2600 |
| PB 6–1 | 0.47±0.05 | 0.32 | H11 to H3 and P12 to P11 | 14000, 1700 |
| PB 6–2 | 0.53±0.03 | 0.36 | H11 to H3 and P12 to P11 | 14000, 1700 |

**Table 5.** Plasma Diagnostic.

| Parameter | Lines | PB8 | NGC 2867–1 | NGC 2867–2 | PB6–1 | PB6–2 |
|---|---|---|---|---|---|---|
| $n_e$ (cm$^{-3}$) | [N I] (λ5198)/(λ5200) | — | 1100±450 | 850±300 | — | — |
| | [O II] (λ3726)/(λ3729) | 2650±750 | 3050±800 | 2300±550 | 2050±600 | 2200±550 |
| | [S II] (λ6716)/(λ6731) | 2450±1000 | 3550±1200 | 2500±800 | 1750±900 | 2450±850 |
| | [Cl III] (λ5518)/(λ5538) | 2400±1800 | 4750±1100 | 4750±1200 | — | — |
| | [Ar IV] (λ4711)/(λ4740) | <6850[c] | 6800:[c] | 3900±1000 | 1650±1000 | 1750±950 |
| | $n_e$ (average) | 2550±550 | 4150±500 | 2850±400 | 1900±450 | 2200±400 |
| $T_e$ (K) | [N II] (λ6548+λ6583)/(λ5755) | 8900±500[a] | 11750±400 | 11750±400 | 12350±950 | 12800±550 |
| | [S II] (λ6716+λ6731)/(λ4069+λ4076) | — | 8450:[d] | 8250:[d] | — | — |
| | [O II] (λ3726+λ3729)/(λ7320+λ7330) | 7050±400[b] | — | — | — | — |
| | [O III] (λ4959+λ5007)/(λ4363) | 6900±150 | 11850±300 | 11600±250 | 15750±600 | 15300±500 |
| | [Ar III] (λ7136+λ7751)/(λ5192) | — | 10800±550 | 11350±850 | — | — |
| | [Ar IV] (λ4711+λ4740)/(λ7170+λ7263) | — | 15400:[c] | >19870: | — | — |
| | $T_e$(NII+OIII) | 7000±150 | 11850±250 | 11600±300 | 15450±500 | 15050±450 |
| | He I | 6250±150 | 10900±250 | 10250±250 | 13250±550 | 13250±450 |
| | Balmer Decrement | $5100^{+1300}_{-900}$ | $8950^{+2900}_{-1900}$ | $8950^{+2900}_{-1900}$ | — | — |

[a] Corrected for recombination contribution to [N II] λ5755 line (see text).
[b] Corrected for recombination contribution to [O II] λλ7320+30 lines (see text).
[c] Ar IV λ4740.17 line affected by charge transfer in the CCD.
[d] S II λλ4068.60, 4076.35 lines corrected for the contributions of C III λ4067.94 line and O II λ4075.86, respectively.

**Table 6.** Atomic data.

| | CELs | | ORLs | |
|---|---|---|---|---|
| Ion | Trans. Probabilities | Coll. strengths | Eff. recomb. coeffs. | Comments |
| C$^{++}$ | — | — | Davey et al. (2000) | Case B |
| C$^{3+}$ | — | — | Pequignot et al. (1991) | Case A |
| C$^{4+}$ | — | — | Pequignot et al. (1991) | Case A |
| N$^{+}$ | Wiese et al. (1996) | Lennon & Burke (1994) | — | |
| N$^{++}$ | — | — | Kisielius & Storey (2002) | Case B; triplets |
| | | | Escalante & Victor (1990) | Case A; singlets |
| N$^{3+}$ | — | — | Pequignot et al. (1991) | Case A |
| O$^{+}$ | Zeippen (1982) | Pradhan et al. (2006) | — | |
| O$^{++}$ | Wiese et al. (1996) | Aggarwal & Keenan (1999) | Storey (1994) | Case B; quartets |
| | Storey & Zeippen (2000) | | Liu et al. (1995) | Case A; doublets |
| O$^{3+}$ | — | — | Pequignot et al. (1991) | Case A |
| Ne$^{++}$ | Baluja & Zeippen (1988) | McLaughlin & Bell (2000) | Kisielius & Storey (unpublished) | Case A; quartets |
| Ne$^{3+}$ | Zeippen (1982) | Giles (1981) | — | |
| Ne$^{4+}$ | Baluja (1985) | Griffin & Badnell (2000) | — | |
| S$^{+}$ | Mendoza & Zeippen (1982) | Keenan et al. (1996) | — | |
| S$^{++}$ | Mendoza & Zeippen (1982) | Tayal & Gupta (1999) | — | |
| Cl$^{++}$ | Mendoza (1982) | Mendoza (1983) | — | |
| Cl$^{3+}$ | Mendoza & Zeippen (1982) | Galavís et al. (1995) | — | |
| | Kaufman & Sugar (1986) | | — | |
| Ar$^{++}$ | Mendoza & Zeippen (1983) | Galavís et al. (1995) | — | |
| Ar$^{3+}$ | Mendoza & Zeippen (1982) | Ramsbottom et al. (1997) | — | |
| Ar$^{4+}$ | Mendoza & Zeippen (1982) | Galavís et al. (1995) | — | |

**Table 7.** $He^+$ and $He^{++}$ abundances.

|  | PB8 | NGC 2867–1 | NGC 2867–2 | PB6–1 | PB6–2 |
|---|---|---|---|---|---|
| Lines | 12 | 12 | 12 | 4 | 4 |
| $He^+/H^{+\,b}$ | 1222±22 | 771±18 | 868±24 | 276±31 | 354±36 |
| $\tau_{3889}$ | 6.55±0.87 | 3.65±1.14 | 5.19±1.25 | 1.48: | 4.10: |
| $\chi^2$ | 12.4 | 16.5 | 11.1 | 0.34 | 1.45 |
| $He^{++}/H^+$ | — | 325±6 | 342±7 | 1614±41 | 1439±33 |

[a] In units of $10^{-4}$. Errors correspond to the uncertainties in line flux measurements.
[b] It includes all uncertainties related with line intensities, $n_e$, $\tau_{3889}$ and $t^2$.

**Table 8.** CELs ionic abundances.

| | \multicolumn{9}{c}{$12 + \log(X^{+i}/H^+)$} | | | | | | | | |
|---|---|---|---|---|---|---|---|---|
| | PB8 | | NGC 2867–1 | | NGC 2867–2 | | PB6–1 | PB6–2 |
| Ion | $t^2 = 0.00$ | $t^2 > 0.00$ | $t^2 = 0.00$ | $t^2 > 0.00$ | $t^2 = 0.00$ | $t^2 > 0.00$ | $t^2 = 0.00$ | $t^2 = 0.00$ |
| $N^+$ | 6.79±0.06 | 6.91±0.06 | 6.85±0.04 | 6.95±0.04 | 6.98±0.05 | 7.12±0.05 | 7.08±0.07 | 7.27±0.04 |
| $O^+$ | 7.34±0.10 | 7.48±0.11 | 7.32±0.07 | 7.43±0.07 | 7.47±0.06 | 7.62±0.07 | 6.93±0.10 | 7.14±0.07 |
| $O^{++}$ | 8.74±0.04 | 9.10±0.08 | 8.43±0.03 | 8.60±0.05 | 8.45±0.03 | 8.70±0.06 | 7.92±0.04 | 8.06±0.04 |
| $Ne^{++}$ | 8.11±0.05 | 8.51±0.09 | 7.77±0.04 | 7.95±0.06 | 7.83±0.04 | 8.10±0.07 | 7.36±0.05 | 7.41±0.16 |
| $Ne^{3+}$ | — | — | 7.18±0.20 | 7.35±0.20 | 7.22±0.21 | 7.48±0.21 | 7.85±0.16 | 7.80±0.16 |
| $Ne^{4+}$ | — | — | 5.26±0.13 | 5.44±0.13 | 5.19±0.14 | 5.47±0.15 | 7.29±0.10 | 7.34±0.10 |
| $S^+$ | 5.02±0.08 | 5.14±0.08 | 5.40±0.07 | 5.50±0.07 | 5.55±0.06 | 5.69±0.07 | 5.13±0.09 | 5.31±0.06 |
| $S^{++}$ | 6.99±0.05 | 7.40±0.12 | 6.35±0.04 | 6.53±0.06 | 6.40±0.04 | 6.68±0.07 | 6.28±0.05 | 6.25±0.05 |
| $Cl^{++}$ | 5.30±0.08 | 5.64±0.10 | 4.78±0.04 | 4.95±0.05 | 4.87±0.04 | 5.11±0.06 | — | — |
| $Cl^{3+}$ | — | — | 4.56±0.14 | 4.69±0.14 | 4.47±0.11 | 4.67±0.12 | 4.71±0.13 | 4.73±0.08 |
| $Ar^{++}$ | 6.61±0.03 | 6.91±0.07 | 5.92±0.04 | 6.06±0.05 | 5.97±0.04 | 6.18±0.06 | 5.71±0.05 | 5.80±0.04 |
| $Ar^{3+}$ | 5.00±0.12 | 5.37±0.14 | 5.72±0.04 | 5.89±0.06 | 5.55±0.04 | 5.46±0.06 | 5.99±0.05 | 5.99±0.04 |
| $Ar^{4+}$ | — | — | 4.41±0.13 | 4.56±0.13 | 4.31±0.19 | 4.42±0.19 | 5.56±0.10 | 5.46±0.10 |

**Table 9.** Ionic abundance ratios from permitted lines in PB 8[a]

| | | O$^{++}$/H$^+$ | |
|---|---|---|---|
| Mult. | $\lambda_0$ | $I(\lambda)/I(H\beta)$ ($\times 10^{-2}$) | X$^{+i}$/H$^+$ ($\times 10^{-5}$) |
| 1[b] | 4638.85 | 0.040±0.010 | 122 |
| | 4641.81 | 0.380±0.030 | 153 |
| | 4649.14 | 0.458±0.037 | 148 |
| | 4650.84 | 0.221±0.027 | 125 |
| | 4661.64 | 0.222±0.027 | 116 |
| | 4673.73 | 0.037: | 113: |
| | 4676.24 | 0.184±0.024 | 210 |
| | Sum | | **141** |
| 2 | 4317.14[c] | 0.210±0.027 | — |
| | 4319.63 | 0.081±0.021 | 112 |
| | 4325.76 | 0.036: | 261: |
| | 4336.83 | 0.054±0.019 | 183 |
| | 4345.56[c] | 0.257±0.028 | — |
| | 4349.43 | 0.197±0.026 | 100 |
| | 4366.89 | 0.199±0.026 | 260 |
| | Sum | | **141** |
| 10[d] | 4069.62 | 0.391±0.035 | 158 |
| | 4069.89 | * | * |
| | 4072.15 | 0.265±0.029 | 109 |
| | 4075.86 | 0.275±0.030 | 79 |
| | 4085.11 | 0.086±0.022 | 191 |
| | 4092.93 | 0.038: | 114: |
| | Sum | | **113** |
| 20[d] | 4097.22[c] | 0.281±0.031 | — |
| | 4110.79 | 0.147±0.025 | 601 |
| | 4119.22 | 0.087±0.022 | 97 |
| | Sum | | **204** |
| 3d–4f[d] | 4089.29 | 0.102±0.022 | 85 |
| | 4303.82[e] | 0.086±0.022 | 174 |
| | 4609.44 | 0.033: | 64: |
| | Average | | **85** |
| | Adopted | | **141±7** |
| | | C$^{++}$/H$^+$ | |
| 6 | 4267.26 | 0.781±0.055 | **69** |
| | Adopted | | **69±5** |
| | | N$^{++}$/H$^+$ | |
| 48 | 4239.40 | 0.128±0.040 | **30** |
| | Adopted | | **30±6** |
| | | Ne$^{++}$/H$^+$ | |
| 57 | 4391.94 | 0.062±0.020 | 15 |
| | 4409.30 | 0.061±0.020 | 24 |
| | Sum | | **19** |
| | Adopted | | **19±6** |

[a] Only lines with intensity uncertainties lower than 40 % have been considered (see text).
[b] Corrected from NLTE effects (see text).
[c] Blended with another line or affected by internal reflections or charge transfer in the CCD.
[d] Recombination coefficients for intermediate coupling (Liu et al., 1995).
[e] Blended with O II $\lambda$4303.61 line.

**Table 10.** Ionic abundance ratios from permitted lines in NGC 2867[a]

| | | NGC2867–1 | | NGC2867–2 | |
|---|---|---|---|---|---|
| Mult. | $\lambda_0$ | $I(\lambda)/I(H\beta)$ $(\times 10^{-2})$ | $X^{+i}/H^+$ $(\times 10^{-5})$ | $I(\lambda)/I(H\beta)$ $(\times 10^{-2})$ | $X^{+i}/H^+$ $(\times 10^{-5})$ |
| | | $O^{++}/H^+$ | | | |
| 1[b] | 4638.85 | 0.040±0.010 | 45 | — | 54 |
| | 4641.81 | 0.034±0.010 | 41 | 0.030±0.015 | 51 |
| | 4649.14 | 0.056±0.011 | 38 | 0.036±0.018 | 56 |
| | 4650.84 | 0.050±0.011 | 42 | 0.030±0.015 | 45 |
| | 4661.64 | 0.038±0.010 | 35 | 0.018±0.014 | 38 |
| | 4676.24 | — | — | — | — |
| | 4696.36 | — | — | — | — |
| | Sum | | **40** | | **50** |
| 2 | 4345.56 | 0.020: | 25: | 0.038: | 48: |
| | 4349.43 | 0.048±0.015 | 24 | 0.040: | 22: |
| | Sum | | 24 | | 29: |
| 10[c] | 4072.15 | 0.122±0.018 | 51 | 0.171±0.032 | 71 |
| | 4075.86 | 0.168±0.025[d] | 51: | 0.236±0.045 | 71: |
| | Sum | | 51 | | 71 |
| | Adopted | | **40±3** | | **50±5** |
| | | $C^{++}/H^+$ | | | |
| 6 | 4267.26 | 0.814±0.049 | **81** | 1.246±0.075 | **123** |
| 17.04 | 6461.95 | 0.074±0.012 | 72 | 0.108±0.021 | 104 |
| | Adopted | | **81±5** | | **123±7** |
| | | $C^{3+}/H^+$ | | | |
| 1 | 4647.42 | 0.342±0.024 | 50 | 0.366±0.037 | 53 |
| | 4650.25 | 0.212±0.019 | 51 | 0.339±0.034 | 83 |
| | 4651.47 | 0.109±0.016 | 79 | 0.097±0.026 | 73 |
| | Sum | | **53** | | **65** |
| 16 | 4067.94 | 0.151±0.020 | 35 | 0.328±0.039 | 76 |
| | 4068.91[e] | 0.198±0.040 | 35 | 0.430±0.086 | 76 |
| | 4070.26 | 0.407±0.028 | 56 | 0.386±0.039 | 53 |
| | Sum | | 44 | | 66 |
| 18 | 4186.90 | 0.227±0.023 | **39** | 0.239±0.033 | **41** |
| | Adopted | | **50±4** | | **65±5** |
| | | $C^{4+}/H^+$ | | | |
| | 4657.15 | 0.083: | 2: | 0.245: | 6: |
| 8.01 | 7726.20 | 0.133±0.013 | 7 | 0.077±0.016 | 4 |
| | Adopted | | **7±1** | | **4±1** |

[a] Only lines with intensity uncertainties lower than 40 % have been considered (see text)
[b] Corrected from NLTE effects (see text).
[c] Recombination coefficients for intermediate coupling (Liu et al., 1995).
[d] Deblended from [S II] $\lambda$4076.35 line, using the theoretical strength ratio $\lambda$4075.86/$\lambda$4072.15.
[e] Deblended from [S II] $\lambda$4068.60 by assuming the theoretical ratio C III $\lambda$4067.94/$\lambda$4068.91.

**Table 11.** Ionic abundance ratios from permitted lines in PB 6[a]

| | | PB 6–1 | | PB 6–2 | |
|---|---|---|---|---|---|
| Mult. | $\lambda_0$ | $I(\lambda)/I(H\beta)$ $(\times 10^{-2})$ | $X^{+i}/H^+$ $(\times 10^{-5})$ | $I(\lambda)/I(H\beta)$ $(\times 10^{-2})$ | $X^{+i}/H^+$ $(\times 10^{-5})$ |
| | | $C^{++}/H^+$ | | | |
| 6 | 4267.26 | 0.487: | 49: | 1.029±0.309 | **105** |
| | Adopted | | **49:** | | **105±31** |
| | | $C^{3+}/H^+$ | | | |
| 1 | 4647.42 | 1.153±0.415 | 154 | 0.636: | 86: |
| | 4650.25 | 0.734: | 163: | 0.553: | 124: |
| | Sum | | 154 | | 100: |
| 18 | 4186.90 | 0.807: | 150: | 0.948±0.303 | 174 |
| | Adopted | | **154±55** | | **174±56** |
| | | $C^{4+}/H^+$ | | | |
| 8.01 | 7726.20 | 0.954±0.191 | 52 | 0.678±0.129 | 37 |
| | Adopted | | **52±10** | | **37±7** |

[a] Only lines with intensity uncertainties lower than 40 % have been considered (see text)

**Table 12.** Total abundances.

| | CELs (12 + log(X/H)) | | | | | | | |
|---|---|---|---|---|---|---|---|---|
| | PB8 | | NGC 2867–1 | | NGC 2867–2 | | PB6–1 | PB6–2 |
| Ion | $t^2 = 0.00$ | $t^2 > 0.00$ | $t^2 = 0.00$ | $t^2 > 0.00$ | $t^2 = 0.00$ | $t^2 > 0.00$ | $t^2 = 0.00$ | $t^2 = 0.00$ |
| He | 11.09±0.01 | 11.09±0.01 | 11.04±0.01 | 11.04±0.01 | 11.08±0.01 | 11.08±0.01 | 11.28±0.01 | 11.25±0.01 |
| O | 8.76±0.04 | 9.11±0.08 | 8.56±0.03 | 8.73±0.04 | 8.59±0.03 | 8.83±0.05 | 8.51±0.04 | 8.58±0.04 |
| N | 8.21±0.11 | 8.55±0.11 | 8.09±0.08 | 8.25±0.09 | 8.10±0.06 | 8.33±0.08 | 8.68±0.09 | 8.71±0.09 |
| Ne | 8.13±0.07 | 8.52±0.12 | 7.87±0.06 | 8.05±0.06 | 7.93±0.06 | 8.20±0.07 | 8.05±0.11 | 8.04±0.11 |
| S | 7.31±0.10 | 7.80±0.10 | 6.63±0.04 | 6.82±0.05 | 6.65±0.04 | 6.94±0.06 | 6.51±0.07 | 6.48±0.09 |
| Cl | 5.30±0.08 | 5.64±0.10 | 5.09±0.09 | 5.24±0.09 | 5.12±0.06 | 5.36±0.08 | 5.58±0.14 | 5.43±0.09 |
| Ar | 6.64±0.04 | 6.93±0.07 | 6.18±0.03 | 6.31±0.04 | 6.14±0.04 | 6.36±0.06 | 6.18±0.05 | 6.22±0.04 |
| | ORLs (12 + log(X/H)) | | | | | | | |
| O[a] | 9.16±0.02 | 9.16±0.02 | 8.73±0.03 | 8.73±0.03 | 8.83±0.04 | 8.83±0.04 | — | — |
| C[b] | 8.85±0.03 | 8.85±0.03 | — | — | — | — | — | — |
| C[c] | — | — | 9.17±0.03 | 9.17±0.03 | 9.32±0.03 | 9.32±0.03 | 9.46: | 9.54±0.10 |

[a] For PB 8, $O^{++}/H^+$ derived from ORLs and $O^+/H^+$ from CELs and $t^2$. For NGC 2867 we have adopted an ICF to correct for unseen ionization stages (see text).
[b] $C/H = ICF(C) \times C^{++}/H^+$.
[c] $C/H = ICF(C) \times (C^{++}/H^+ + C^{3+}/H^+ + C^{4+}/H^+)$.

**Table 13.** $t^2$ parameter.

| Method | PB8 | NGC 2867–1 | NGC 2867–2 | PB6–1 | PB6–2 |
|---|---|---|---|---|---|
| $O^{++}$ (R/C) | 0.037±0.005 | 0.045±0.009 | 0.061±0.010 | — | — |
| $Ne^{++}$ (R/C) | 0.016±0.012 | — | — | — | — |
| $He^+$ | $0.008^{+0.020}_{-0.028}$ | 0.039±0.019 | 0.053±0.022 | 0.102:: | 0.139:: |
| Bac–CELs | $0.038^{+0.013}_{-0.016}$ | $0.061^{+0.041}_{-0.061}$ | $0.056^{+0.050}_{-0.056}$ | — | — |
| **Adopted** | **0.033±0.005** | **0.044±0.008** | **0.060±0.010** | — | — |